\documentclass[prd,onecolumn,superscriptaddress,floatfix,longbibliography]{revtex4}

\usepackage{amsfonts}
\usepackage{amsmath}
\usepackage{graphicx}
\usepackage{amssymb}
\usepackage{color}
\usepackage{subfigure}
\usepackage{mathtools}
\usepackage{tikz-cd}
\usepackage[colorlinks]{hyperref}
\usepackage[normalem]{ulem}

\newcommand{\ket}[1]{|{#1}\rangle}
\newcommand{\bra}[1]{\langle{#1}|}

\global\long\def\ket#1{\left|#1\right\rangle }
\global\long\def\bra#1{\left\langle #1\right|}

\begin{document}
\title{Influence functional for two mirrors interacting via radiation pressure}

\author{Salvatore Butera}
\affiliation{School of Physics and Astronomy, University of Glasgow, Glasgow G12 8QQ, United Kingdom}


\begin{abstract}
We study the effective dynamics of two mirrors, forming an optical cavity, and interacting with the cavity field via radiation pressure. We pursue a perturbative influence functional approach to trace out the degrees-of-freedom of the field, and obtain the second order effective action for the system composed by the mirrors. We find that the interaction between the mirrors is mediated by pairs of field modes, which combine in such a way to give rise to two different interaction channels. We find that the quantum and thermal fluctuations of the cavity field result in coloured, Gaussian stochastic noises acting on the mirrors. To each of these noises is associated a dissipative effect, and the corresponding power spectra and susceptibilities are related via generalized fluctuation-dissipation relations. We finally demonstrate that the dynamics of the mirrors admits a stochastic interpretation, and give the relative quantum Langevin equations.
\end{abstract}
\maketitle

\section{Introduction}

It is well known that the vacuum state of a quantum field is not truly empty, but is permeated by fundamental, zero-point fluctuations. The physics that originates from such quantum fluctuations is rich, and underlies a plethora of physical phenomena, such as the Lamb shift of atomic levels, the Casimir and Casimir-Polder forces, and the dynamical Casimir effect \cite{milonni-book}. In optomechanics \citep{Aspelmeyer_RMP}, which is the research field that studies the interaction between macroscopic objects and the electromagnetic field, it has been shown that the vacuum fluctuations can affect mechanical motion \cite{KardarRMP1999} by inducing dissipation and decoherence \cite{Dalvit-PRL-2000,MaiaNeto-PRA-2000,Savasta-PRX-2018,Butera-PRA-2019,Butera-EPL-2019}, and even transfer mechanical energy between physically separated mirrors, as an ordinary fluid \cite{Savasta-PRL-2019}.

Given the recent advances in miniaturization techniques \citep{Aspelmeyer_RMP}, which allow mechanical devices to operate deep in the quantum regime \cite{OConnell-Science-2010,Aspelmeyer-Science-2020}, these findings open up a wide range of technological applications. Devices in which forces between mechanical components are exchanged via the mediation of vacuum fluctuations could find application in those research fields where high sensitivity is required, such as in the search for gravitational waves \cite{Abramovici-science-1992} and dark matter \cite{Carney-QTech-2021}. They could be used in quantum metrology and quantum sensing \cite{QSens-RMP-2017} 
and, in the near future, could serve as actuators of nanoscale motion, for manipulating objects as small as nanowires, quantum dots, or even living organisms such as viruses and bacteria in biological applications \cite{Stange-PhysToday-2021}. Recent proposals have also shown that nano-metric system could be used to realize mechanically based quantum bit \cite{MechQbit-PRX-2021} for quantum information and quantum computing.

Building on these premises, the accurate theoretical modelling of such devices is a highly desirable goal. By moving in this direction, the aim of this paper is to develop a detailed description for the effective dynamics of two vibrating mirrors in a typical optomechanical cavity. The objective is to develop a model that goes beyond the standard single mode approximation, and takes into account the multi-mode nature of the field instead \cite{Kanu-Thesis}.
To this aim, we make use of an open quantum system approach based on the theory of the influence functionals \cite{Feynman-IF,Feynman-book-PathInt}, according to which we identify the field with the environment, while the mirrors as the system of interest. This approach is similar, but pursues the opposite objective, to the one developed in \cite{Mazzitelli-IF-DCE}, where the authors were interested in developing a description of the dynamical Casimir effect that included dissipation and noise from first principles.

We consider the one-dimensional configuration for simplicity, and model the interaction between the mirrors and the field in terms of the radiation pressure \cite{Law-MirFieldInt-1995,Law-MirFieldInt-2011}. Within this framework, the mirrors are assumed perfectly reflecting. This approximation is valid as long as we consider taking part to the interaction only the modes of the field with frequency below the plasma frequency of the materials. More general models, that explicitly account for the internal dynamics of the mirrors have been developed \cite{Law-MirFieldInt-2011,MOF1,MOF2}, and the role played by the internal degrees-of-freedom (dofs) in mediating the interaction between the optical field and the mechanical oscillators have been studied \cite{Unruh-PRD-2014,Kanu-PRD-2021}. The theory developed in this paper is not only relevant to optomechanical systems, but is also applicable for modelling the dynamics of microwave optomechanical circuits \cite{Heikki-PRL-2014} and superconducting circuits \cite{Johansson-PRL-2009,Johansson-PRA-2010,Butera-PRA-2019,Wilson-DCE-Analog-2011}, which are characterized by an equivalent radiation-pressure coupling.

The first original result of this paper generalizes the description of the radiation pressure interaction, developed in \cite{Law-MirFieldInt-1995}, to the case of two dynamical mirrors. Given the nonlinear nature of the resulting coupling between the mechanical and optical dofs, we work in the weak interaction limit and pursue a perturbative generating functional approach \cite{QBM2,Yang-AnPhys-2020} to derive a second order effective action for the system composed by the two mirrors. We find that the quantum and thermal fluctuations of the field appear in the form of coloured noises acting on the mirrors, whose non-equilibrium dynamics is non-Markovian. Interestingly, the structure of the noise and dissipation kernels that characterize this dynamics manifestly shows that couples of field modes are involved in mediating the interaction between the mirrors. This property results from the underlying dynamical Casimir mechanism. We obtain that such kernels are related via generalized fluctuation-dissipation relation \cite{QBM2,JT-PRD-2020}. We finally show that this dynamics admits a stochastic description, and give the corresponding quantum Langevin equations \cite{Calzetta_PhysA}.

The paper is structured as follows: We start by introducing in Sec.~\ref{Sec:System} the physical system at hand and by generalizing the theory of the radiation pressure interaction to the case of two dynamical mirrors. The details of the calculations are reported in Appendix~\ref{App1}. In Sec.~\ref{Sec:EffActMaster} we begin the study of the influence functional for the mirrors. In particular, in Sec.~\ref{Sec:IFdef} we introduce the concept of influence functional and influence action, while in Sec.~\ref{Sec:IFpert} we present their perturbative expansion. The relevant field correlators, and the generating functional from which they are derived are reported in Sec.~\ref{Sec:GF}. In Sec.~\ref{Sec:IA} we present the main results of this paper, that is the second order influence action for the two mirrors, and discuss the physical interpretation of our findings. The details of the calculations leading to these results are reported in Appendix~\ref{App2}. We show in Sec.~\ref{Sec:FlucDiss} that the noise and dissipation kernels that describe the non-equilibrium dynamics of the mirrors satisfy generalized fluctuation-dissipation relations. In Sec.~\ref{Sec:Stoch} we illustrate the stochastic interpretation of the effective dynamics of the mirrors, and give the corresponding quantum Langevin equations. Finally, in Sec.~\ref{Sec:Concl}, we draw our conclusions.


 



\section{The physical system\label{Sec:System}}

\begin{figure}[t]
    \centering
    \includegraphics[width = 3.5 in]{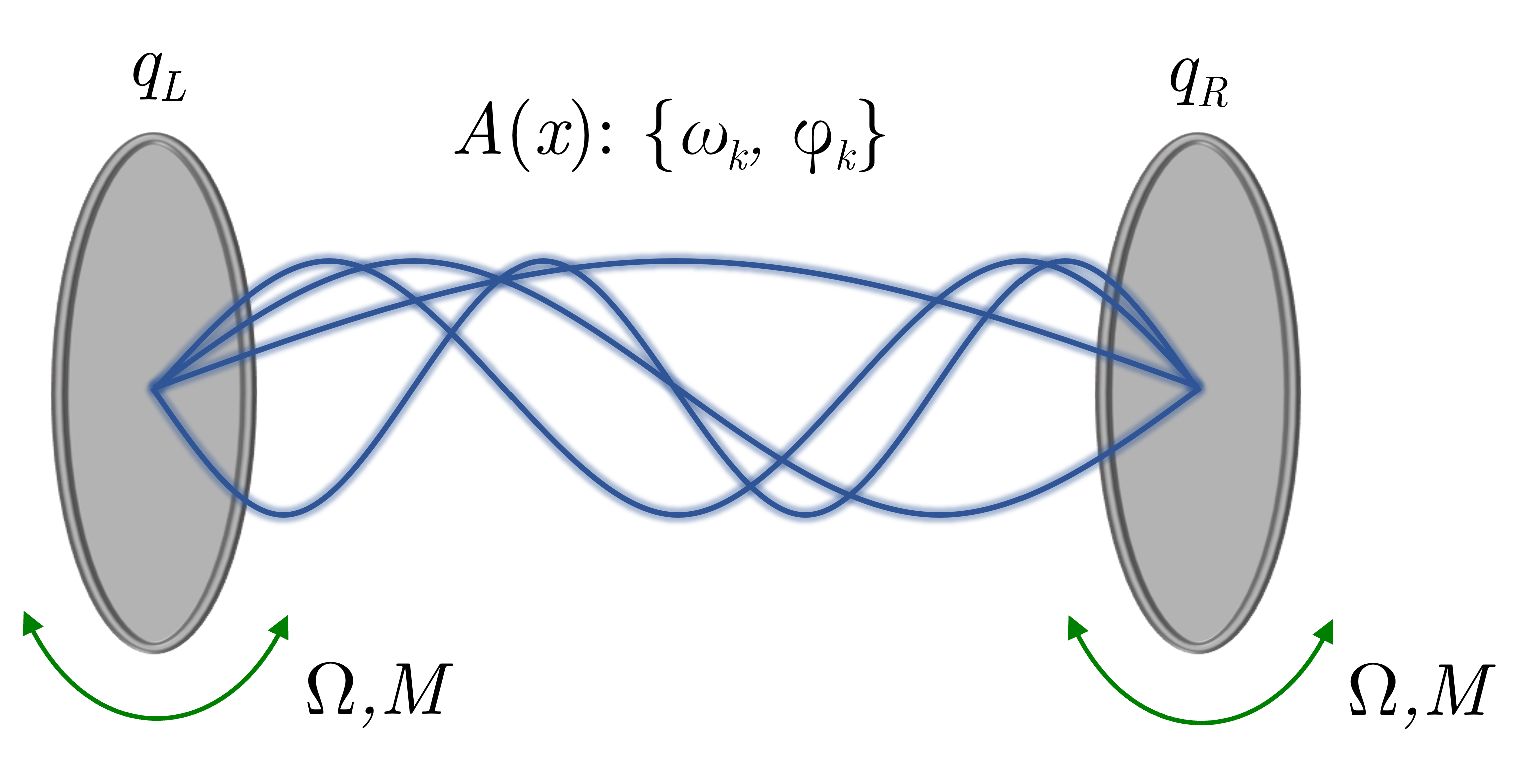}
    \caption{Schematic representation of two mirrors forming an optical cavity, and confined in harmonic potentials of frequency $\Omega$. The mirrors interact with the scalar field $A(x)$ enclosed in the cavity. The coordinates $q_L$ and $ q_R$ correspond to the positions of the left and right mirrors, while $L = q_R - q_L$ is the length of the cavity. We indicate by $ \varphi_k(x)$ the eigenfunctions that give the spatial structure of the cavity modes, and with $\omega_k$ the corresponding frequencies.}
\label{Fig:sch}
\end{figure}

We consider the minimal one-dimensional system composed of two moving mirrors of equal mass $M$, forming an optical cavity, each confined in a harmonic potential of frequency $\Omega$ (see Fig.\ref{Fig:sch}). We indicate respectively with $q_L(t)$ and $q_R(t)$ the positions of the left and right mirrors, and with $q_{L,0}$ and $q_{R,0}$ their value in correspondence of the minimum of the confining potentials. The length of the cavity in such condition is: $L_0 \equiv q_{R,0} - q_{L,0}$. In the region inside the cavity, $q_L(t)\leq x \leq q_R(t)$, is defined the scalar field $A(x)$. This is the equivalent of the vector potential in the case of the electromagnetic field. We make the same assumption as in \cite{Law-MirFieldInt-1995, Law-MirFieldInt-2011} and neglect the effects of the field outside the cavity for simplicity. This procedure is physically justified if there is an appreciable number of photons in the cavity, so that the inside field dominates over the outside one. In this paper, we are however interested in laying down the basis for the treatment of the problem, identifying the fundamental physics, rather than give the precise solution to the equations of motion of the mirrors. The more general case, in which the dofs of the outside field are taken into account, represents a straightforward extension of the results illustrated in this work. This problem will be the object of future further investigation. The mirrors and the field interact with each other via radiation pressure, that we model by imposing that the field vanishes at the mirrors positions: $A(x=q_L(t),t) = A(x=q_R(t),t) = 0$. In other terms, this means working in the limit of divergent electric susceptibility, and thus assuming the mirrors being perfectly reflecting \cite{Law-MirFieldInt-2011,MOF1,MOF2}. This assumption is reliable only for optical modes of frequency below the plasma frequency of the mirrors, which thus provides a natural cut-off for the modes that take part to the interaction. 

Basing on these assumptions, the action $S[A,q_n]$ of the whole system can be written as the sum of the action $S_M[q_n]$, that describes the free dynamics of the mirrors, and the action $S_A'[A,q_n]$, that accounts for the field dofs instead. We used here the shorthand $q_n \equiv \{q_R, q_L\}$, to collectively indicate the dofs of the right and left mirrors. These actions take the explicit form $(c=1)$:
\begin{align}
	&S_M[q_n] = 
	\frac{M}{2}\int_{t_i}^{t_f} dt\sum_{n=R,L}\left[\dot{q}_n^2-\Omega^2(q_n-q_{n,0})^2\right],\label{Eq:LM1}\\
	&S_A'[A,q_n] = 
	\frac{1}{2}\int_{t_i}^{t_f} dt\int_{q_L(t)}^{q_R(t)}{dx \left[(\partial_t A)^2-(\partial_x A)^2\right]},\label{Eq:LA1}
\end{align}
with $t_i$ and $t_f$ respectively the initial and final time of the motion. Note that we omit for brevity the explicit time-dependence of variables in the equations, when not necessary. The action $S_A'[A,q_n]$ contains the dofs of the mirrors, since the spatial integral is extended over the (time-dependent) cavity length: $L(t) \equiv q_R(t) - q_L(t)$.

We proceed by projecting the field onto the instantaneous basis \cite{Law-MirFieldInt-1995,Law-MirFieldInt-2011}, that is composed by the set of eigenfunctions $\{\varphi_k[x;q_n(t)]\}$ $(k = 1,\,2,\,...)$ which vanish in correspondence to the location of the mirrors. These have the form:
\begin{equation}
	\varphi_k[x;q_n(t)] \equiv \sqrt{\frac{2}{q_R(t)-q_L(t)}}\,\sin\left\{\omega_k(t) [x-q_L(t)]\right\},
	\label{Eq:Modes}
\end{equation}
where $\omega_k(t) \equiv k\pi/[q_R(t)-q_L(t)]$ are the cavity-length-dependent frequencies of the modes. Note that the eigenmodes $\varphi_k[x;q_n(t)]$ implicitly depend on time because of the motion of the mirrors. Indicating collectively by $Q_{\{k\}}(t)\equiv \left\{Q_{1},\, Q_{2},\, ...\right\}$ the set of time-dependent amplitudes of the modes, the field is written in this basis as
\begin{equation}
	A(x,t) = \sum_k {Q_k(t)\varphi_k[x;q_n(t)]}.
	\label{Eq:ModesDecomp}
\end{equation}

The action of the field, Eq.~\eqref{Eq:LA1}, can be decomposed in the instantaneous basis by using Eqs.~\eqref{Eq:Modes} and \eqref{Eq:ModesDecomp}. This yields (see Appendix \ref{App1} for further details):
\begin{equation}
	S_A'[Q_{\{k\}},q_n] = \int_{t_i}^{t_f}{dt\,\left\{\frac{1}{2}\sum_k{[\dot{Q}_k^2 - \omega_k^2(t)Q_k^2]} + \sum_{k,j}\frac{\dot{Q}_k Q_j}{q_R-q_L}\dot{x}_{jk} + \sum_{k,j,\ell}{\frac{Q_k Q_j}{(q_R-q_L)^2} \dot{x}_{j\ell}\dot{x}_{k\ell}}\right\}}.
	\label{Eq:LA}
\end{equation}
We use the standard notation and indicate time derivatives with dots over variables. In Eq.~\eqref{Eq:LA}, we defined $\dot{x}_{jk}\equiv\left(r_{jk}\dot{q}_R + l_{jk} \dot{q}_L\right)$, with
\begin{subequations}
\begin{align}
	r_{kj} &= (q_R - q_L)\int_{q_L}^{q_R}{dx\,\frac{\partial\varphi_k}{\partial q_R}\varphi_j} = - (-1)^{k+j} g_{kj},\label{Eq:r_kj}\\
	l_{kj} &= (q_R - q_L)\int_{q_L}^{q_R}{dx\,\frac{\partial\varphi_k}{\partial q_L}\varphi_j} = g_{kj},\label{Eq:l_kj}\\
	g_{kj} &= \frac{2 j k}{j^2-k^2}, \label{Eq:g_kj}
\end{align}
\end{subequations}
and used the following relations, that result from the completeness of the basis $\{\varphi_k[x;q(t)]\}$:
\begin{subequations}
\begin{align}
	\sum_\ell r_{k\ell}r_{j\ell} &= (q_R - q_L)^2\int_{q_L}^{q_R}{dx\,\frac{\partial\varphi_k}{\partial q_R}\frac{\partial\varphi_j}{\partial q_R}},\label{Eq:r_kl-r_jl}\\
	\sum_\ell l_{k\ell}l_{j\ell}&= (q_R - q_L)^2\int_{q_L}^{q_R}{dx\,\frac{\partial\varphi_k}{\partial q_L}\frac{\partial\varphi_j}{\partial q_L}},\label{Eq:l_kl-l_jl}\\
\sum_\ell l_{k\ell}r_{j\ell}&= (q_R - q_L)^2\int_{q_L}^{q_R}{dx\,\frac{\partial\varphi_j}{\partial q_L}\frac{\partial\varphi_k}{\partial q_R}}.\label{Eq:l_kl-r_jl}
\end{align}
\end{subequations}
The action in Eq.~\eqref{Eq:LA} is the first original result of this paper. It extends the theory developed in \cite{Law-MirFieldInt-1995} to the case of two moving mirrors interacting with the field inside the cavity via radiation pressure. The first term in Eq.~\eqref{Eq:LA} is a collection of actions of harmonic oscillators, one for each mode of the field, with time-dependent frequencies. The second and third terms involve the velocities of both the mirrors and the field modes, and account for non-adiabatic effects of their interaction. 

In what follows, we derive an effective action for the system composed by the two mirrors, by using an open quantum system formalism based on the theory of the influence functionals. Due to the nonlinearity of the mirrors-field interaction, that is encoded in Eq.~\eqref{Eq:LA}, we follow a perturbative approach, in which this interaction is treated as a small perturbation to the free dynamics of the mirrors and the field. In preparation to these following steps, it is thus convenient to rewrite Eq.~\eqref{Eq:LA} as:
\begin{equation}
	S_A'[Q_{\{k\}},q_n(t)] = S_A[Q_{\{k\}}] + S_{\rm int}[Q_{\{k\}},q_n],
	\label{Eq:LA2}
\end{equation}
where
\begin{align}
    & S_{A}[Q_{\{k\}}] = \int_{t_i}^{t_f}{dt\,\frac{1}{2}\sum_k{[\dot{Q}_k^2 - \omega_{k,0}^2Q_k^2]}},\label{Eq:SA0}\\
    & S_{\rm int}[Q_{\{k\}},q_n] = \int_{t_i}^{t_f}{dt\,\left\{\frac{1}{2}\sum_k{[\omega_{k,0}^2 - \omega_k^2(t)]Q_k^2} + \sum_{k,j}\frac{\dot{Q}_k Q_j}{q_R-q_L}\dot{x}_{jk} + \sum_{k,j,\ell}{\frac{Q_k Q_j}{(q_R-q_L)^2} \dot{x}_{j\ell}\dot{x}_{k\ell}}\right\}},\label{Eq:Sint}
\end{align}
are respectively the action for the free field, and the action that accounts for the coupling between the optical and mechanical dofs. We indicated here by $\omega_{k,0}\equiv k\pi/L_0$ the frequency of the $k-$th mode, corresponding to the length of the cavity at equilibrium. Gathering these results together, we can write the action for the whole system in the form:
\begin{equation}
    S[Q_{\{k\}},q_n] = S_M[q_n] + S_{A}[Q_{\{k\}}] + S_{\rm int}[Q_{\{k\}},q_n].
    \label{Eq:Action_forPert}
\end{equation}
Eq.~\eqref{Eq:Action_forPert}, will be the starting point for the perturbation theory developed in the following sections.


\section{Effective action for the mirrors\label{Sec:EffActMaster}}
The nonlinear interaction between the mirrors and the field, that is described by the action $S_{\rm int}[Q_{\{k\}},q_n]$ introduced above, prevents us to obtain an exact analytical solution for the effective dynamics of the two mirrors. We therefore assume the strength of this interaction weak enough, and follow a perturbative influence functional approach, as prescribed in \cite{QBM2}. We start in Sec.~\ref{Sec:IFdef} by giving an overview on the influence functional formalism, and by introducing in Sec.~\ref{Sec:IFpert} its perturbative expansion. After defining the generating functional and the relevant Feynman propagators in Sec.~\ref{Sec:GF}, we give in Sec.~\ref{Sec:IA} the explicit expression of the second order action for the system of the two mirrors, and discuss the physics it encodes. We show in Sec.~\ref{Sec:FlucDiss} that the noise and dissipation kernels, which describe the non-Markovian evolution of the system, satisfy generalized fluctuation-dissipation relations. Finally, in Sec.~\ref{Sec:EffAct}, we summarize our results.

\subsection{Definition of influence functional\label{Sec:IFdef}}
The quantum state of the two mirrors can be described in terms of the reduced density matrix $\hat{\rho}_r$, that is obtained by tracing out the dofs of the field from the density matrix $\hat{\rho}$ of the whole system: $\hat{\rho}_r \equiv {\rm Tr}_{\rm A} [\hat{\rho}]$. By working in the position representation, and using the bra-ket Dirac notation, this is written as:
\begin{equation}
	\hat{\rho}_r(t) = \int_{-\infty}^{+\infty} d q_n\int_{-\infty}^{+\infty} d q_n' \,\rho_r( q_n, q_n';t) \ket{ q_n}\bra{ q_n'}.
\end{equation}
The reduced density matrix $\hat{\rho}_r$ is evolved in time via the reduced evolution operator $ J_r( q_{n,f}, q_{n,f}',t_f| q_{n,i}, q_{n,i}',t_i) $, that gives the amplitude of the transition of the system from the initial configuration $ \{q_{n,i},  q_{n,i}'\}$ at the time $t_i$, to the final configuration $ \{q_{n,f},  q_{n,f}'\}$ at the time $t_f$. Accordingly, we write the reduced density matrix at the final time as:
\begin{equation}
	\rho_r( q_n, q_n';t_f) = \int_{-\infty}^{+\infty} d q_{n,i}\int_{-\infty}^{+\infty} d q_{n,i}' \,J_r( q_{n,f}, q_{n,f}',t_f| q_{n,i}, q_{n,i}',t_i) \rho_r( q_{n,i},  q_{n,i}';t_i).
\end{equation}
We assume that the mirrors and the field are uncorrelated at the initial time, such that the density matrix for the whole system factorizes as $\hat{\rho}(t=0) = \hat{\rho}_r\otimes\hat{\rho}_{\rm A}$, where $\hat{\rho}_{\rm A}$ is the density matrix for the field sector. Given this assumption, and by using the path integral formalism \cite{Feynman-IF,Feynman-book-PathInt}, the reduced evolution operator can be written in the form: 
\begin{equation}
 J_r( q_{n,f}, q_{n,f}',t_f| q_{n,i}, q_{n,i}',t_i) = \int_{ q_{n,i} , t_i}^{ q_{n,f} , t_f}\mathcal{D} q_n\int_{ q_{n,i}' , t_i}^{ q_{n,f}' , t_f}\mathcal{D} q_n' \exp\left[\frac{i}{\hbar}(S_M[ q_n]-S_M[ q_n'])\right]\,F[ q_n,  q_n'],
 \label{Eq:Prop1}
\end{equation}
where
\begin{multline}
	F[ q_n,  q_n'] \equiv \int_{-\infty}^{+\infty} dQ_{\{k\},f} \int_{-\infty}^{+\infty} dQ_{\{k\},i} \int_{-\infty}^{+\infty} dQ_{\{k\},i}'\;\rho_{\rm A}(Q_{\{k\},i},Q_{\{k\},i}';t_i)\\
	\times\left\{\int_{Q_{\{k\},i},t_i}^{Q_{\{k\},f},t_f}\mathcal{D}Q_{\{k\}}\int_{Q_{\{k\},i}',t_i}^{Q_{\{k\},f},t_f}\mathcal{D}Q_{\{k\}}'
	\exp\left[\frac{i}{\hbar}\left(S_{A}[Q_{\{k\}}] + S_{\rm int}[ Q_{\{k\}},q_n]  - S_{A}[Q_{\{k\}}'] - S_{\rm int}[ Q_{\{k\}}',q_n]\right)\right]\right\}
	\label{Eq:IF_def}
\end{multline}
is the formal expression for the influence functional. This encodes the effect of the field onto the dynamics of the two mirrors. For readability, we introduced in Eq.~\eqref{Eq:IF_def} the notation of the type: $dQ_{\{k\}} \equiv \prod_k{dQ_k}$ and $\mathcal{D}Q_{\{k\}} \equiv \prod_k{\mathcal{D}Q_k}$, respectively for standard integrals over the field dofs, and for integrals over the relative paths. The influence functional in Eq.~\eqref{Eq:IF_def} can equivalently be written in terms of the influence action $\delta A [ q_n, q_n']$, that is defined as:
\begin{equation}
 	F[ q_n,  q_n'] \equiv \exp\left(\frac{i}{\hbar} \delta A[ q_n,  q_n']\right).
 	\label{Eq:IA_def}
\end{equation}
By using Eqs.~\eqref{Eq:IF_def} and \eqref{Eq:IA_def}, we can write the evolution operator Eq.~\eqref{Eq:Prop1} in terms of the effective action $S_{M}^{\rm eff}[ q_n,  q_n']\equiv S_M[ q_n]-S_M[ q_n']+\delta A[ q_n,  q_n']$, as:
\begin{equation}
 J_r( q_{n,f}, q_{n,f}',t_f| q_{n,i}, q_{n,i}',t_i) = \int_{ q_{n,i} , t_i}^{ q_{n,f} , t_f}\mathcal{D} q_n\int_{ q_{n,i}' , t_i}^{ q_{n,f}' , t_f}\mathcal{D} q_n' \exp\left(\frac{i}{\hbar}S_{M}^{\rm eff}[ q_n,  q_n']\right).
 \label{Eq:Prop2}
\end{equation}
From this results it is thus clear that the effect of the field onto the dynamics of the system composed by the mirrors is accounted for by the influence action, whose evaluation is the key task in the forthcoming sections.

\subsection{Perturbative expansion of the influence action\label{Sec:IFpert}}
Due to the nonlinearity of the mirrors-field interaction $S_{\rm int}[Q_{\{k\}},q_n]$, an exact solution for the influence action cannot be obtained. To overcome this difficulty, we pursue a perturbative approach that treats such an interaction as a perturbation to the free dynamics of the mirrors and the field. The perturbative expansion of the influence functional (or influence action) is obtained by noticing that the definition in Eq.~\eqref{Eq:IF_def} can be interpreted as an average over the dofs of the field. By introducing the average of the generic function $\mathcal{O}\big(Q_{\{k\}},Q'_{\{k\}}\big)$:
\begin{multline}
\big<\mathcal{O}\big(Q_{\{k\}},Q'_{\{k\}}\big)\big>_{\rm A} \equiv \int_{-\infty}^{\infty} dQ_{\{k\},f} \int_{-\infty}^{\infty} dQ_{\{k\},i} \int_{-\infty}^{\infty} dQ_{\{k\},i}' \;\rho_{\rm A}(Q_{\{k\},i},Q_{\{k\},i}',t_i)\\
	\times\left\{\int_{Q_{i},t_i}^{Q_{f},t_f}\mathcal{D}Q_{\{k\}}\int_{Q'_{i},t_i}^{Q'_{f},t_f}\mathcal{D}Q_{\{k\}}'e^{\frac{i}{\hbar}\left(S_A[Q_{\{k\}}]-S_A[\{Q_k'\}]\right)}\mathcal{O}\big(Q_{\{k\}},Q'_{\{k\}}\big)\right\},
\label{Eq:AverageDef}
\end{multline}
the influence functional can be written as
\begin{equation}
 	F[ q_n, q_n'] = 
 	\left< \exp\left[\frac{i}{\hbar}\left(S_{\rm int}[ Q_{\{k\}},q_n]  - S_{\rm int}[ Q_{\{k\}}',q_n']\right)\right] \right>_{\rm A}.
 	\label{Eq:InflFuncAv}
\end{equation}
The sought perturbative expansion is thus obtained by power expanding the exponential in Eq.~\eqref{Eq:InflFuncAv}. Up to second order in $S_{\rm int}[Q_{\{k\}},q_n]$, the corresponding influence action is written as:
\begin{equation}
	\delta A[ q_n, q_n'] \approx \delta A^{(1)}[ q_n, q_n'] + \delta A^{(2)}[ q_n, q_n'],
	\label{Eq:IA1}
\end{equation}
where
\begin{align}
&\begin{aligned}
	\delta A^{(1)}[ q_n,  q'_n] &= \big<S_{\rm int}[Q_{\{k\}},q_n]\big>_{\rm A} - \big<S_{\rm int}[Q'_{\{k\}},q'_n]\big>_{\rm A},\label{Eq:A1}
\end{aligned}\\
&\begin{aligned}
	\delta A^{(2)}[ q_n, q'_n ] &= \frac{i}{2\hbar}\left\{ \big<{S_{\rm int}}^2[ Q_{\{k\}},q_n]\big>_{\rm A} - \big<S_{\rm int}[Q_{\{k\}},q_n]\big>_{\rm A}^2 \right\}
		+\frac{i}{2\hbar}\left\{ \big<{S_{\rm int}}^2[ Q'_{\{k\}},q'_n]\big>_{\rm A} - \big<S_{\rm int}[ Q'_{\{k\}},q'_n]\big>_{\rm A}^2 \right\}\\
		&-\frac{i}{\hbar}\left\{\big<S_{\rm int}[ Q_{\{k\}},q_n]S_{\rm int}[ Q'_{\{k\}},q'_n]\big>_{\rm A} - \big<S_{\rm int}[ Q_{\{k\}},q_n]\big>_{\rm A}\big<S_{\rm int}[Q'_{\{k\}}, q'_n]\big>_{\rm A}\right\},\label{Eq:A2}
\end{aligned}
\end{align}
are the first- and second-order contributions, respectively. Given Eq.~\eqref{Eq:Sint}, Eqs.~\eqref{Eq:A1} and \eqref{Eq:A2} are defined respectively in terms of second- and fourth-order correlators of the field that involve both the amplitudes of the field modes $Q_{\{k\}}$ and their velocities $\dot{Q}_{\{k\}}$. These correlators can be obtained from the generating functional $\mathcal{G}[J_{\{k\}},J_{\{k\}}']$ of the free field that we introduce in the following section.

\subsection{Generating functional and the propagators\label{Sec:GF}}

\noindent The correlators involving only the amplitudes $Q_{\{k\}}$ of the field modes are obtained by differentiating the corresponding generating functional, that is defined as
\begin{equation}
	\mathcal{G}[J_{\{k\}},J_{\{k\}}'] \equiv \left<\exp\left\{\frac{i}{\hbar}\sum_k\int_{t_i}^{t_f} dt\left[J_k(t)Q_k(t)-J_k'(t)Q_k'(t)\right]\right\}\right>_{\rm A}.
	\label{Eq:GF}
\end{equation}
This formally represents the influence functional of the field, in the case its dofs are linearly coupled to the external actions $J_{\{k\}}$. Averages of the type in Eq.~\eqref{Eq:AverageDef} can be computed by simply differentiating the generating functional, that is:
\begin{equation}
	\big<\mathcal{O}[Q_{\{k\}},Q_{\{k\}}']\big>_{\rm A} = \mathcal{O}\bigg[\frac{\hbar}{i}\frac{\delta}{\delta J_{\{k\}}},-\frac{\hbar}{i}\frac{\delta}{\delta J_{\{k\}}'}\bigg]\mathcal{G}\big[ J_{\{k\}},J_{\{k\}}'\big]|_{J_{\{k\}}=J_{\{k\}}'=0}.
	\label{Eq:Average}
\end{equation}
The explicit expression for Eq.~\eqref{Eq:GF} can be calculated by using standard path integral techniques \cite{QBM3}. By  considering an initial thermal state for the field, with temperature $T$, this takes the form \cite{QBM1,QBM2,QBM3}
\begin{widetext}
\begin{multline}
	\mathcal{G}[J_{\{k\}},J_{\{k\}}'] = \exp\left\{-\frac{i}{\hbar}\sum_k\int_{t_i}^{t_f} ds_1\int_{t_i}^{s_1} ds_2\left[J_k(s_1) - J_k'(s_1)\right]\mu_{k}(s_1-s_2)\left[J_k(s_2) + J_k'(s_2)\right] \right.\\
	\left.-\sum_k\frac{1}{\hbar}\int_{t_i}^{t_f} ds_1\int_{t_i}^{s_1} ds_2\left[J_k(s_1) - J_k'(s_1)\right]\nu_{k}(s_1-s_2)\left[J_k(s_2) - J_k'(s_2)\right] \right\},
\end{multline}
where the dissipation $\mu_k$ and noise $\nu_k$ kernels are defined as
\begin{subequations}
\begin{align}
	\nu_{k}(t) &= \frac{z_k}{2\omega_{k,0}}\cos(\omega_{k,0} t),\label{Eq:Nu}\\
	\mu_{k}(t) &= -\frac{1}{2\omega_{k,0}}\sin(\omega_{k,0} t),\label{Eq:Mu}
\end{align}
\end{subequations}
and $z_k \equiv \coth\left({\hbar\omega_{k,0}}/{2k_B T}\right)$. These are related via the fluctuation-dissipation relation \cite{QBM1,QBM3}:
\begin{equation}
	\nu_k(t) = z_k\omega_{k,0}\gamma_k(t),
\label{Eq:FDR-Field}
\end{equation}
where $\gamma_k(t) \equiv \int^t{ds\, \mu_k(s)}=[1/(2\omega_{k,0}^2)]\cos(\omega_{k,0} t)$. Since the action of the free field is quadratic, averages of order higher than two can be decomposed as product of averages of order equal or lower than two. First-order averages are identically zero, as well as the second-order cross-correlations between the different, independent modes of the field. The following Feynman propagators exhaust all the relevant second-order averages involving the field amplitudes:
\begin{subequations}
\begin{align}
	\mathcal{D}_{Q_k Q_k}(s_1-s_2) \equiv \left< Q_k(s_1) Q_k(s_2) \right> = -i\hbar[&-\mu_{k}(s_1-s_2){\rm sgn}(s_1-s_2) +i\nu_{k}(s_1-s_2)],\\
	\mathcal{D}_{Q'_k Q'_k}(s_1-s_2) \equiv \left< Q_k'(s_1) Q_k'(s_2) \right> = -i\hbar[&\mu_{k}(s_1-s_2){\rm sgn}(s_1-s_2) + i\nu_{k}(s_1-s_2)],\\
	\mathcal{D}_{Q'_k Q_k}(s_1-s_2) \equiv \left< Q_k'(s_1) Q_k(s_2) \right> = -i\hbar[&\mu_{k}(s_1-s_2)- i\nu_{k}(s_1-s_2)].
\end{align}
\label{Eq:PropQ}
\label{Eq:Propagators}
\end{subequations}
Correlators involving the velocities $\dot{Q}_k$ are readily obtained by taking appropriate time-derivatives of Eqs.~\eqref{Eq:Propagators}(a-c) (see Appendix \ref{App2}).
\end{widetext}

\subsection{Influence action for the mirrors\label{Sec:IA}}
We arrive now to the main result of this paper, that is the second-order influence action for the mirrors. We leave the details of the calculations to Appendix \ref{App2}. Here we focus on presenting the main results and discussing their physical meaning. We work in the limit of small oscillations of the mirrors around their equilibrium positions $q_{n,0}$ within the confining potentials, and write: $q_n(t) = q_{n,0} + \delta q_{n}(t)$, where $\delta q_n(t)$ denotes fluctuations of the mirrors' positions. In other words, we assume the length scale of the mechanical motion much smaller than the wavelength of optical modes that interact with the mirrors (that is, ultimately, the wavelength corresponding to the plasma frequency). We present the following results up to the second-order in the perturbative parameter $\delta q_n/L_0$.

The first order term $\delta A^{(1)}[ q_n,  q'_n]$ of the influence action involves only the variation $\delta q_\Delta = \delta q_R-\delta q_L$ in the relative distance between the mirrors. Because of this reason, we find convenient to indicate it as $A_\Delta^{(1)}[\delta q_\Delta,\delta q_\Delta']\equiv\delta A^{(1)}[ q_n,  q'_n]$. This is composed by two terms:
\begin{align}
	A_\Delta^{(1)}[\delta q_\Delta] =\delta A_{\Delta,\delta}^{(1)}[\delta q_\Delta,\delta q_\Delta'] + \delta A_{\Delta,\delta^2}^{(1)}[\delta q_\Delta,\delta q_\Delta'].
	\label{Eq:A1_12}
\end{align}
The former, $\delta A_{\Delta,\delta}^{(1)}$, is linear in the fluctuations $\delta q_\Delta$ and accounts for the Casimir force that attracts the two mirrors. The latter, $\delta A_{\Delta,\delta^2}^{(1)}$, is instead quadratic in $\delta q_\Delta$ and is responsible for a correction to the frequency of the harmonic potential that limit the physical separation between the two mirrors. These are static effects that arise because of the cavity-length-dependence of the frequencies of the modes of the field. The bare contribute of these terms formally diverges, but they can be renormalized by subtracting the corresponding values in the limit of infinite separation between the mirrors, as discussed in Appendix \ref{App2}. Note that, by following this procedure, we make use of the energy of the field outside of the cavity, in order to compensate the infinite change of the vacuum energy inside the cavity, that is consequent to the change in the mutual distance between the mirrors. This entails taking into account the static contribute of the interaction between the mirrors and the outside field, while its dynamics is neglected. As noted in \cite{Law-MirFieldInt-1995}, strictly speaking, this approach is not fully self-consistent, as also the dynamical effects of such an interaction should be taken into account. This can however be achieved by simply generalizing our approach and consider also the dofs of the field outside the cavity. Nevertheless, we remark that neglecting the dynamics of the outside field is a good approximation in those physical configurations dominated by the field inside the cavity (i.e. in those cases in which an appreciable number of photons are in the cavity).

The two first-order terms introduced above take a simple form in the zero temperature limit:
\begin{subequations}
\begin{align}
	\delta A_{\Delta,\delta}^{(1)}[\delta q_\Delta,\delta q_\Delta'] &	=  - \int_{t_i}^{t_f}{dt \,\left(\frac{\hbar\pi}{24L_0^2}\right)\left(\delta q_\Delta - \delta q_\Delta'\right)},\qquad\;\;\;\; \text{low temp.}\\
\delta A_{\Delta,\delta^2}^{(1)}[\delta q_\Delta,\delta q_\Delta'] &=  \int_{t_i}^{t_f}{dt \,\left(\frac{3\hbar\pi}{16L_0^3}\right)\left[(\delta q_\Delta^2) - (\delta q_\Delta')^2\right]},\qquad \text{low temp.}
\end{align}
\end{subequations}
as well as in the high temperature limit:
\begin{subequations}
\begin{align}
	\delta A_{\Delta,\delta}^{(1)}[\delta q_\Delta,\delta q'_\Delta] 	&=  \int_{t_i}^{t_f}{dt \,\left(\frac{k_B T}{2L_0}\right)\left(\delta q_\Delta - \delta q_\Delta'\right)},\qquad\qquad\;\;\;\;\;\;\;\, \text{high temp.}\\
	 \delta A_{\Delta,\delta^2}^{(1)}[\delta q_\Delta,\delta q'_\Delta] &=  - \int_{t_i}^{t_f}{dt \,\left(\frac{3k_B T}{4L_0^2}\right)\left[(\delta q_\Delta^2) - (\delta q_\Delta')^2\right]}. \qquad\;\;\;\; \text{high temp.}
\end{align}
\end{subequations}

The second order term of the influence action, $\delta A^{(2)}[\delta q_n, \delta q'_n]$, describes instead the dynamical effects of the coupling between the mirrors. It can be decomposed into two different contributions: one, we call it $\delta A_\Delta^{(2)}[\delta q_\Delta, \delta q'_\Delta]$, describes the dynamical evolution of the mutual distance $\delta q_\Delta = \delta q_R - \delta q_L$ between the mirrors; the other, $\delta A_\Sigma^{(2)}[\delta q_\Sigma,\delta q'_\Sigma]$, accounts instead for the dynamics of their center-of-mass (CM) $\delta q_\Sigma = \delta q_R + \delta q_L$. These take the explicit form:
\begin{align}
&\begin{aligned}
     \delta A_{\Delta}^{(2)}[\delta q_\Delta,\delta q_\Delta'] &= \hbar\left\{\int_{t_i}^{t_f} ds_1 \int_{t_i}^{s_1} ds_2  \left[-i\delta{q}_\Delta^{(-)}(s_1)\bar{N}(s_1-s_2)\delta{q}_\Delta^{(-)}(s_2)+\delta{q}_\Delta^{(-)}(s_1)\bar{M}(s_1-s_2)\delta{q}_\Delta^{(+)}(s_2)\right]\right\}\\
        &+\hbar\Bigg\{\int_{t_i}^{t_f}ds_1 \int_{t_i}^{s_1} ds_2\Bigg[-i \delta\dot{q}_{\Delta}^{(-)}(s_1)N_\Delta(s_1 - s_2)\delta\dot{q}_{\Delta}^{(-)}(s_2)+\delta\dot{q}_{\Delta}^{(-)}(s_1)M_\Delta(s_1 - s_2)\delta\dot{q}_{\Delta}^{(+)}(s_2)\Bigg]\Bigg\},\label{Eq:A_Delta}
\end{aligned}\\[18pt]
&\begin{aligned} 
            \delta A_{\Sigma}^{(2)}[\delta q_\Sigma,\delta q_\Sigma'] &= \hbar\Bigg\{\int_{t_i}^{t_f}ds_1 \int_{t_i}^{s_1} ds_2\Bigg[-i \delta\dot{q}_{\Sigma}^{(-)}(s_1)N_\Sigma(s_1 - s_2)\delta\dot{q}_{\Sigma}^{(-)}(s_2)+\delta\dot{q}_{\Sigma}^{(-)}(s_1)M_\Sigma(s_1 - s_2)\delta\dot{q}_{\Sigma}^{(+)}(s_2)\Bigg].\label{Eq:A_Sigma}
\end{aligned}
\end{align}
In Eqs.~\eqref{Eq:A_Delta} and \eqref{Eq:A_Sigma}, we defined the combination of the forward- and backward-in-time histories $\delta q_i^{(\pm)} = (\delta q_i \pm \delta q_i')$ (with $i=\Delta,\,\Sigma$). This result shows that the mirrors follow a non-Markovian, out-of-equilibrium dynamics. We will demonstrate in Sec.~\ref{Sec:Stoch} that the imaginary components in the influence action physically represent noises acting onto the mirrors, which are due to the quantum and thermal fluctuations of the cavity field. The real terms represent instead the corresponding dissipation counterparts.

The term in the first bracket in Eq.~\eqref{Eq:A_Delta} is a result of the cavity-length-dependence of the frequencies of the field modes. It accounts for a non-local in time coupling for the relative distance between the mirrors, and is defined in therms of the noise $\bar{N}(t)$ and dissipation $\bar{M}(t)$ kernels. These have the form:
\begin{align}
    \bar{N}(t) & = \sum_k \frac{\omega_{k,0}^2}{4L_0^2}\bar{N}_k(t), &	\bar{M}(t) & = \sum_k \frac{\omega_{k,0}^2}{4L_0^2}\bar{M}_k(t),\label{Eq:Kernels_bar_main}
\end{align}
with 
\begin{align}
	\bar{N}_k(t) & = \nu_+(t;k,k) + \nu_-(t;k,k), &
\bar{M}_k(t) & = \mu_+(t;k,k),\label{Eq:NMbar}
\end{align}
and
\begin{align}
 \nu_\pm(t;k,j) &\equiv  - (z_k z_j \pm 1) \cos[(\omega_k\pm\omega_j)t], &
 \mu_\pm(t;k,j) &\equiv   (z_k \pm z_j ) \sin[(\omega_k\pm\omega_j)t].\label{Eq:nm_pm}
\end{align}
Eqs.~\eqref{Eq:NMbar} and \eqref{Eq:nm_pm} show that this interaction is mediated via two different channels, that we call respectively the $(+)$ and $(-)$ channels. To these correspond the noise and dissipation kernels $\nu_+(t;k,k)$, $\mu_+(t;k,k)$ and $\nu_-(t;k,k)$, $\mu_-(t;k,k)\equiv 0$, respectively. Notice here the peculiar behaviour of the $(-)$ coupling channel: in this case the noise kernel $\nu_-(t;k,k)$ vanishes in the zero-temperature limit, and it has no dissipation counterpart at any temperature.

The term in the second bracket in Eq.~\eqref{Eq:A_Delta}, as well as Eq.~\eqref{Eq:A_Sigma}, account instead for a coupling between the velocities of the two mirrors. In Eq.~\eqref{Eq:A_Delta} this contribute is appreciable respect to the first term, which involves the positions, in the case the frequency $\Omega$ of the mechanical vibrations is of the same order of the optical frequencies $\omega_{\{k\}}$. By considering state-of-the-art microwave resonators, this regime can be achieved with hybrid quantum electromechanical system \cite{Rouxinol-2016} or ultra-high-frequency micromechanical resonators \cite{Cleland2010}. Also in this case, this coupling is non-local in time, and the corresponding kernels take the form:
\begin{align}
    N_{\Delta}(t) &= {\sum_{\substack{kj}}}''\frac{\omega_k\omega_j}{4L_0^2} N_{kj}(t), & M_{\Delta}(t) &= {\sum_{\substack{kj}}}''\frac{\omega_{k,0}\omega_{j,0}}{4L_0^2} M_{kj}(t),\label{Eq:Kernels_Delta}\\
    N_{\Sigma}(t) &= {\sum_{\substack{kj}}}'\frac{\omega_k\omega_j}{4L_0^2} N_{kj}(t),  & M_{\Sigma}(t)&= {\sum_{\substack{kj}}}'\frac{\omega_{k,0}\omega_{j,0}}{4L_0^2} M_{kj}(t).\label{Eq:Kernels_Sigma}
\end{align}
with
\begin{align}
	N_{k,j}(t) &= N_{k,j}^{(+)}(s) + N_{k,j}^{(-)}(t), &
	M_{k,j}(t) &= M_{k,j}^{(+)}(s) + M_{k,j}^{(-)}(t),\label{Eq:NM_kj}
\end{align}
and
\begin{subequations}
\begin{align}
	N_{k,j}^{(\pm)}(t) &= \frac{1}{(\omega_{k,0}\pm\omega_{j,0})^2} \nu_\pm(t;k,j), & 
	M_{k,j}^{(\pm)}(t) &= \frac{1}{(\omega_{k,0}\pm\omega_{j,0})^2} \mu_\pm(t;k,j).\label{Eq:NM_kj_pm}
\end{align}
\end{subequations}
In Eqs.~\eqref{Eq:Kernels_Delta}, double primed sums indicate sums over couples of cavity modes $k,j$, such that $k+j=2n$ $(n \in \mathbb{N})$ is an even number while, in Eq.~\eqref{Eq:Kernels_Sigma}, single primed sums indicate sums over modes such that $k+j=2n+1$ is an odd number instead. This shows that the relative distance and the CM dofs of the mirrors interact with an environment that is composed by different combinations of field modes. It is interesting to note that the structure of the noise and dissipation kernels defined above reveals that mechanical energy is exchanged between the mirrors via the mediation of pairs of field modes, which is the hallmark of the underlying dynamical Casimir mechanism. According to this picture, the fluctuations in one of the mirrors decay in favour of the creation of pairs of photons inside the cavity. These photons travel towards the other end of the cavity, where they interact and thus excite the second mirror \cite{Savasta-PRL-2019}.
Also in this case, the modes that mediate the interaction between the mirrors combine in such a way to give rise to two different coupling channels whose noise and dissipation kernels are respectively $N_{k,j}^{(+)}(t)$, $M_{k,j}^{(+)}(t)$ and $N_{k,j}^{(-)}(t)$, $M_{k,j}^{(-)}(t)$.

Note that the noise and dissipation kernels in Eqs.~\eqref{Eq:Kernels_bar_main}, \eqref{Eq:Kernels_Delta} and \eqref{Eq:Kernels_Sigma} diverge and cannot be renormalized by using the field outside the cavity, as done for the static, first order terms. This is a limit of the radiation pressure model \cite{Giulio-PRL-2013,Armata-PRD-2015,Bartolo2015,Armata-PRD-2017}. Such a limit is however overcome by physical considerations, that are related to the transparency of the mirrors for modes of frequency beyond the plasma frequency of the materials. In other words, in the sums that define the kernels, a cut-off $\Gamma$ of the order of the plasma frequency, needs to be introduced.


\subsection{Fluctuation-dissipation relations\label{Sec:FlucDiss}}
The noise and dissipation kernels we obtained from the nonlinear mirrors-field coupling at hand, are related via generalized fluctuations-dissipation relations. By defining $\gamma_\pm({t;k,j}) \equiv \int^t{ds\,\mu_\pm(s;k,j)}$, these can be easily expressed in the frequency domain by noting that:
\begin{align}
   \nu_+(t;k,j) &= (\omega_{k,0}+\omega_{j,0}) \frac{z_k z_j +1}{z_k + z_j} \gamma_+(t;k,j),\label{FDR+}\\
   \nu_-(t;k,j) &= (\omega_{k,0} - \omega_{j,0}) \frac{z_k z_j - 1}{z_j - z_k} \gamma_-(t;k,j).\label{FDR-}
\end{align}
These differ from the standard fluctuation-dissipation relation, Eq.~\eqref{Eq:FDR-Field}, which characterize a linear system-environment coupling. They are however identical to the linear ones, both in the high temperature and zero temperature limits. In the former case, $k_B T/\hbar\Gamma \gg 1$, we have:
\begin{equation*}
    z_k \to 2K_B T/\hbar\omega_{k,0},    
\end{equation*}
and the fluctuation-dissipation relation takes the standard Kubo relation \cite{Feynman-IF,Caldeira-PhysA-1983,QBM2}:
\begin{equation}
    \nu_\pm(t;k,j) = \frac{2 k_B T}{\hbar} \gamma_\pm(t;k,j).
\end{equation}
In the opposite zero-temperature limit, $z_k \to 1$, and we obtain instead:
\begin{align}
    \nu_+(t;k,j) &= (\omega_{k,0}+\omega_{j,0}) \gamma_+(t;k,j),\label{Eq:FDR_T0+}\\
    \nu_-(t;k,j) &= |\omega_{k,0}-\omega_{j,0}| \gamma_-(t;k,j).\label{Eq:FDR_T0-}
\end{align}
This is a general property of the fluctuation-dissipation relations \cite{QBM2}. Physically, it means that these relations are insensitive to the type of system-environment coupling both at high and low temperature. Eqs.~\eqref{Eq:FDR_T0+} and \eqref{Eq:FDR_T0-} highlight once again that the interaction between the two mirrors is mediated by pairs of field modes, which combine in such a way to give rise to two different coupling channels. In the $(+)$ channel, the effective frequencies of the bath modes the mirrors interact with, is given by the sum of the frequencies of the modes that take part to the interaction. In the $(-)$ channel instead, this is given by their difference. As a final remark notice that, since the constant noise term $\nu_-(t;k,k)$ in Eq.~\eqref{Eq:Kernels_bar_main} has no dissipation counterpart, there is no way to form a fluctuation-dissipation relation in this case. Such a term vanishes in the zero-temperature limit, and describes a delta-correlated, white noise acting on the cavity length dof of the mirrors.

\subsection{Effective actions\label{Sec:EffAct}}
We collect here the results presented in the previous sections, and write the effective action for the system composed by the two mirrors. We have found that, up to second-order in the perturbation theory, the influence action decouples into the two independent contributes for the CM and the relative distance dofs (notice that a coupling is expected to appear at higher orders). We find thus convenient to write the action for the free mirrors, Eq.~\eqref{Eq:LM1}, in the same basis. These take the form:
\begin{equation}
	S_M[\delta q_n] = S_{\Sigma,0} [\delta q_\Sigma] + S_{\Delta,0} [\delta q_\Delta],
\end{equation}
with
\begin{subequations}
\begin{align}
	S_{\Sigma,0} [\delta q_\Sigma] &= \int_{t_i}^{t_f}{dt\left[\frac{M}{4}\left(\delta\dot{q}_\Sigma^2-\Omega^2\delta q_\Sigma^2\right)\right]},\label{Eq:S_sigma0}\\
	S_{\Delta,0} [\delta q_\Delta] &= \int_{t_i}^{t_f}{dt\left[\frac{M}{4}\left(\delta\dot{q}_\Delta^2-\Omega^2\delta q_\Delta^2\right)\right]}.\label{Eq:S_delta0}
\end{align}
\end{subequations}
By inserting the free actions $S_{\Sigma,0} [\delta q_\Sigma]$ and $S_{\Delta,0} [\delta q_\Delta]$, together with the first $A_\Delta^{(1)}[\delta q_\Delta]$ and the second order contributes $A_\Delta^{(2)}[\delta q_\Delta]$, $A_\Sigma^{(2)}[\delta q_\Sigma]$ of the influence action, into the effective action $S_{M}^{\rm eff}[ q_n,  q_n']$ that defines the reduced propagator in Eq.~\eqref{Eq:Prop2}, we can write
\begin{equation}
	S_M^{\rm eff} = S_{\Sigma,{\rm eff}} + S_{\Delta,{\rm eff}},
\end{equation}
where
\begin{subequations}
\begin{align}
	S_{\Sigma,{\rm eff}} &= S_{\Sigma,0}[\delta q_\Sigma]-S_{\Sigma,0}[\delta q_\Sigma']+A_\Sigma^{(2)}[\delta q_\Sigma],\label{Eq:Seff_s}\\
    S_{\Delta,{\rm eff}} &= S_{\Delta,0}[\delta q_\Delta]-S_{\Delta,0}[\delta q_\Delta']+A_\Delta^{(1)}[\delta q_\Delta]+A_\Delta^{(2)}[\delta q_\Delta], \label{Eq:Seff_d}
\end{align}
\end{subequations}
are the effective actions for the CM and the mutual distance dofs, respectively. The reduced propagator can be factorized accordingly:
\begin{equation}
 J_r(\delta q_{n,f},\delta q_{n,f}',t_f|\delta q_{n,i},\delta q_{n,i}',t_i) =  J_{r,\Sigma}(\delta q_{\Sigma,f},\delta q_{\Sigma,f}',t_f|\delta q_{\Sigma,i},\delta q_{\Sigma,i}',t_i)\times  J_{r,\Delta}(\delta q_{\Delta,f},\delta q_{\Delta,f}',t_f|\delta q_{\Delta,i},\delta q_{\Delta,i}',t_i),
\end{equation}
with:
\begin{subequations}
\begin{align}
    J_{r,\Sigma}(\delta q_{\Sigma,f},\delta q_{\Sigma,f}',t_f|\delta q_{\Sigma,i},\delta q_{\Sigma,i}',t_i) &= \int_{\delta q_{\Sigma,i} , t_i}^{\delta q_{\Sigma,f} , t_f}\mathcal{D}\delta q_\Sigma\int_{\delta q_{\Sigma,i}' , t_i}^{\delta q_{\Sigma,f}' , t_f}\mathcal{D}\delta q_\Sigma' \exp\left[i S_{\Sigma,{\rm eff}}/\hbar\right],\\
	J_{r,\Delta}(\delta q_{\Delta,f},\delta q_{\Delta,f}',t_f|\delta q_{\Delta,i},\delta q_{\Delta,i}',t_i) &= \int_{\delta q_{\Delta,i} , t_i}^{\delta q_{\Delta,f} , t_f}\mathcal{D}\delta q_\Delta\int_{\delta q_{\Delta,i}' , t_i}^{\delta q_{\Delta,f}' , t_f}\mathcal{D}\delta q_\Delta' \exp\left[i S_{\Delta,{\rm eff}}/\hbar\right].
\end{align}
\end{subequations}

\section{Stochastic interpretation of the influence functional\label{Sec:Stoch}}
We show in this section that the theory of the influence functional previously discussed allows for a stochastic description of the effective dynamics of the mirrors \cite{Calzetta_PhysA}. In particular, we show that the imaginary part of the influence action, which comprises the noise kernels, can be interpreted as originating from coloured, Gaussian stochastic noises acting on the mirrors. For definiteness, we develop the following arguments by considering the influence action in Eq.~\eqref{Eq:A_Delta}, since it is the most general as both the position-position and velocity-velocity coupling appear. For simplicity, we discuss separately the stochastic models for these two types of interactions.


\subsubsection{Position-position coupling}
The dynamics introduced by the first term in Eq.~\eqref{Eq:A_Delta} can be interpreted in term of a stochastic force $\xi_\Delta(s)$ coupled to the position $\delta q_\Delta$. To prove this, let us consider the following action:
\begin{equation}
    S_\xi \left[\delta q_\Delta\right]= \int_{t_i}^{t_f} {dt\left\{L[\delta q_\Delta(t),\delta\dot{q}_\Delta(t)] + \delta q_\Delta(t) \xi_\Delta(t)\right\}},
    \label{Eq:S_cl_1}
\end{equation}
in which $\xi_\Delta(t)$ is a Gaussian stochastic process, with non-zero mean. The effect of the force $\xi_\Delta$ onto the dynamics of the system can be described in terms of its corresponding influence functional. Indicating with $P[\xi(t)]$ the probability density functional for the $\xi_\Delta(t)$ histories, this takes the form:
\begin{equation}
    F_{\xi}\left[\delta q_\Delta,\delta q_\Delta'\right] = \int{\mathcal{D}\xi_\Delta(t) P[\xi_\Delta(t)] \exp\left\{\frac{i}{\hbar}\left[\delta q_\Delta(t)-\delta q_\Delta'(t)\right]\xi_\Delta(t)\right\}}.
\label{Eq:IF_xi}
\end{equation}
This equation formally defines the characteristic functional of the stochastic process $\xi_\Delta(t)$ \cite{gardiner-book-qNoise}. Since $\xi_\Delta(t)$ is Gaussian by hypothesis, only the first two cumulants $C_{\xi,1}$ and $C_{\xi,2}$ are different from zero. These are defined as:
\begin{equation}
    C_{\xi,1}(t) \equiv \left<\xi_\Delta(t)\right>, \qquad C_{\xi,2}(t_1,t_2) \equiv \left<\xi_\Delta(t_1)\xi_\Delta(t_2)\right> - \left<\xi_\Delta(t_1)\right>\left<\xi_\Delta(t_2)\right>.
\end{equation}
The influence functional in Eq.~\eqref{Eq:IF_xi} can be thus written in the form \cite{Feynman-book-PathInt}:
\begin{equation}
    F_\xi = \exp\left\{\frac{i}{\hbar} \int_{t_i}^{t_f}{dt\, C_{\xi,1}(t)\delta q_\Delta^{(-)}(t)}-\frac{1}{2\hbar^2}\int_{t_i}^{t_f}ds_1\int_{t_i}^{t_f} ds_2 \delta q_\Delta^{(-)}(s_1) C_{\xi,2}(s_1,s_2) \delta q_\Delta^{(-)}(s_2)\right\},
\label{Eq:IF_xi_2}
\end{equation}
while the probability density functional reads
\begin{equation}
    P[\xi_\Delta(t)] = \exp\left\{ -\frac{1}{2}\int_{t_i}^{t_f}ds_1\int_{t_i}^{t_f}ds_2 \left[\xi_\Delta(s_1) - C_{\xi,1}(s_1)\right] \left[C_{\xi,2}\right]^{-1}(s_1,s_2) \left[\xi_\Delta(s_2) - C_{\xi,1}(s_2)\right]\right\}.
    \label{Eq:PDF_Pos}
\end{equation}
By comparing Eq.~\eqref{Eq:IF_xi_2} with the first term in Eq.~\eqref{Eq:A_Delta}, we deduce that the two influence functionals have the same structure, given
\begin{subequations}
\begin{align}
    &C_{\xi,1}(t) = \hbar\int_{t_i}^{t}ds\Big[{\bar{M}(t-s) \delta q_\Delta^{(+)}(s)}\Big],\\
    &C_{\xi,2}(t_1,t_2) = -\hbar^2\bar{N}(t_1-t_2).
\end{align}
\end{subequations}
This result provides the sought connection between the quantum and the stochastic pictures. In what follows, we find convenient to separate the average (deterministic) evolution of $\xi_\Delta(t)$ from its fluctuating component. That is, we write $\xi_\Delta(t) \equiv C_{\xi,1}(t)+\tilde{\xi}_\Delta(t)$, where $\tilde{\xi}_\Delta(t)$ is the corresponding zero-mean stochastic process.

\subsubsection{Velocity-velocity coupling}
A similar discussion can be developed also for the second term in Eq.~\eqref{Eq:A_Delta}, that involves the coupling between the velocities of the mirrors. We show here that this term can be interpreted in terms of a non-zero mean, Gaussian stochastic action $\eta_\Delta(s)$, that is coupled to the velocities of the mirrors instead. We follow the same procedure as before and introduce the action:
\begin{equation}
    S_{\eta}\left[\delta q_\Delta\right] = \int_{t_i}^{t_f} {dt\left\{L[\delta q_\Delta(t),\delta\dot{q}_\Delta(t)] + \delta \dot{q}_\Delta(t) \eta_\Delta(t)\right\}}.
    \label{Eq:S_cl_2}
\end{equation}
The effect of the coupling of the velocity with the stochastic process $\eta_\Delta(t)$ is described by the influence functional:
\begin{equation}
    F_{\eta} \left[\delta q_\Delta,\delta q_\Delta'\right] = \int{\mathcal{D}\eta_\Delta(t) P[\eta_\Delta(t)] \exp\left\{\frac{i}{\hbar} \left[\delta\dot{q}_\Delta(t)-\delta\dot{q}_\Delta'(t)\right] \eta_\Delta(t)\right\}}.
    \label{Eq:IF_eta_1}
\end{equation}
Again, Eq.~\eqref{Eq:IF_eta_1} defines the characteristic functional of the stochastic process $\eta_\Delta(t)$. By pursuing analogous arguments as detailed in the previous section, we can write this in terms of the first and second order cumulants of the process $\eta_\Delta(t)$:
\begin{equation}
    C_{\eta,1}(t) = \left<\eta_\Delta(t)\right>, \qquad C_{\eta,2}(t_1,t_2) = \left<\eta_\Delta(t_1)\eta_\Delta(t_2)\right> - \left<\eta_\Delta(t_1)\right>\left<\eta_\Delta(t_2)\right>,
\end{equation}
as
\begin{equation}
    F_\eta = \exp\left\{\frac{i}{\hbar} \int_{t_i}^{t_f}{dt\, C_{\eta,1}(t)\delta \dot{q}^{(-)}(t)}-\frac{1}{2\hbar^2}\int_{t_i}^{t_f}ds_1\int_{t_i}^{t_f} ds_2 \delta \dot{q}^{(-)}(s_1) C_{\eta,2}(s_1,s_2) \delta \dot{q}^{(-)}(s_2)\right\}.
    \label{Eq:IF_eta_2}
\end{equation}
The probability density functional $P[\eta_\Delta(t)]$ takes a form analogous to Eq.~\eqref{Eq:PDF_Pos}. By comparing Eq.~\eqref{Eq:IF_eta_2} with the second term in \eqref{Eq:A_Delta}, we deduce that the two influence functionals have the same structure, given:
\begin{subequations}
\begin{align}
    &C_{\eta,1}(t) = \hbar \int_{t_i}^{t}ds\Big[{M_\Delta(t-s) \delta \dot{q}^{(+)}(s)}\Big],\\
    &C_{\eta,2}(t_1,t_2) = -\hbar^2 N_\Delta(t_1-t_2). \label{Eq:C_eta2}
\end{align}
\end{subequations}
Again, this result provides the connection between the quantum and the stochastic descriptions. We find convenient also in this case to separate the average evolution of $\eta_\Delta$ from its fluctuating component. That is, we write $\eta_\Delta(t) \equiv C_{\eta,1}(t)+\tilde{\eta}_\Delta(t)$, where $\tilde{\eta}_\Delta(t)$ is the corresponding zero-mean stochastic process.

\section{Langevin equations}
The results of the previous section allow us to interpret the imaginary terms in Eqs.~\eqref{Eq:Seff_s} and \eqref{Eq:Seff_d}, as stochastic noises coupled either with the positions or the velocities of the mirrors. According to this picture, we can write the effective actions for the CM and the relative distance dofs of the mirrors as:
\begin{align}
&\begin{aligned}
	S_{\Sigma,{\rm eff}} &= S_{\Sigma,0}[\delta q_\Sigma]-S_{\Sigma,0}[\delta q_\Sigma']+{\rm Re}\{A_\Sigma^{(2)}[\delta q_\Sigma]\}+\int_{t_i}^{t_f} dt\left[\delta \dot{q}_\Sigma(t) - \delta \dot{q}_\Sigma'(t)\right]\tilde{\eta}_\Sigma(t),\label{Eq:Seff_s2}
\end{aligned}\\
&\begin{aligned}
    S_{\Delta,{\rm eff}} &= S_{\Delta,0}[\delta q_\Delta]-S_{\Delta,0}[\delta q_\Delta']+A_\Delta^{(1)}[\delta q_\Delta]+{\rm Re}\{A_\Delta^{(2)}[\delta q_\Delta]\}+\int_{t_i}^{t_f}dt\left[\delta q_\Delta(t) - \delta q_\Delta'(t)\right]\tilde{\xi}_\Delta(t)\\
    &+\int_{t_i}^{t_f}dt\left[\delta \dot{q}_\Delta(t) - \delta \dot{q}_\Delta'(t)\right]\tilde{\eta}_\Delta(t).\label{Eq:Seff_d2}
\end{aligned}
\end{align}
In Eq.~\eqref{Eq:Seff_s2}, we defined the zero-mean Gaussian noise $\tilde{\eta}_\Sigma(s)$, whose second order  correlation function is defined as in Eq.~\eqref{Eq:C_eta2}, with $N_\Sigma$ in place of $N_\Delta$. By extremizing Eq.~\eqref{Eq:Seff_s2} and Eq.~\eqref{Eq:Seff_d2} with respect to $\delta q_\Sigma^{(-)}$ and $\delta q_\Delta^{(-)}$, that is by posing $\partial S_{\Sigma,{\rm eff}}/\partial \delta q_\Sigma^{(-)} = 0$ and $\partial S_{\Delta,{\rm eff}}/\partial \delta q_\Delta^{(-)} = 0$, we obtain the associated quantum Langevin equations of motion for $\delta q_\Sigma^{(+)}$ and $\delta q_\Delta^{(+)}$, respectively. These have the form:
\begin{align}
&\begin{aligned}
   & \frac{M}{4}\delta\ddot{q}_\Delta^{(+)}(t) + \Bigg(\frac{M}{4}\Omega^2 - \frac{3\hbar\pi}{16L_0^3}\Bigg)\delta q_\Delta^{(+)}(t)+ \hbar \int_{t_i}^{t}{ds\Big[\dot{M}_\Delta(t-s) \delta \dot{q}_\Delta^{(+)}(s)\Big]}\\
   &\hspace{40mm}- \hbar\int_{t_i}^{t}{ds\Big[\bar{M}(t-s) \delta q_\Delta^{(+)}(s)\Big]} =- \frac{\hbar\pi}{24 L_0^2} +\dot{\tilde{\eta}}_\Delta(t)-\tilde{\xi}(t),\label{Eq:LangD}
\end{aligned}\\
&\begin{aligned}
   & \frac{M}{2}\delta\ddot{q}_\Sigma^{(+)}(t) + \frac{M}{2}\Omega^2\delta q_\Sigma^{(+)}(t) + \hbar\int_{t_i}^{t}ds\Big[{\dot{M}_\Sigma(t-s) \delta \dot{q}_\Sigma^{(+)}(s)}\Big]  = \dot{\tilde{\eta}}_\Sigma(t).\label{Eq:LangS}
\end{aligned}
\end{align}
It has been shown \cite{Calzetta_PhysA} that these equations can be used to propagate in time the Wigner distribution of the two mirrors, which provides a semi-classical representation for their quantum state \cite{Milburn-Walls}. Moreover, quantum mechanical correlators of operators that are symmetrically ordered respect their hermitian conjugates, can be calculated by performing both an average over the quantum fluctuations that characterize the initial state of the mirrors and a stochastic average respect to the different realizations of the processes $\delta q_\Sigma^{(+)}(t)$ and $\delta q_\Delta^{(+)}(t)$, that are solutions of Eqs.~\eqref{Eq:LangD} and \eqref{Eq:LangS}. We refer the interested readers to Ref.~\cite{Calzetta_PhysA}, for a deeper insight onto this matter. Up to the second order in the perturbation theory, these equations thus provide a stochastic picture for the full quantum dynamics of the system. Notice that an equivalent description for such a dynamics can be formulated \cite{QBM1,QBM2,QBM3}, in terms of a master equation for the reduced density operator of the mirrors.

\section{Conclusions\label{Sec:Concl}}
We have studied the effective dynamics of two mirrors forming an optical cavity, and interacting with the field via radiation pressure. We pursued this objective by following an open quantum system strategy based on the theory of the influence functionals. Due to the nonlinear nature of the coupling, we used a perturbative approach to trace over the degrees-of-freedom of the field and obtain a second order effective action for the system composed by the two mirrors. We found that the mirrors interact with each other via the mediation of pairs of modes of the field. In particular, we showed that each of these pairs mediate the interaction by two disjoint channels. In one of these channels, the frequency of the effective bath mode that results coupled to the system is given by the sum of the frequencies of the two field modes that make up the pair. In the other channel instead, such a frequency is equal to the difference between the frequencies of the modes. The mirrors are coupled both via their positions and velocities, and undergo a non-Markovian evolution. We showed that the quantum and thermal fluctuations of the field induce noises acting on the mirrors. To each of these noises is associated a dissipative effect, and the corresponding memory kernels are related to each other via generalized fluctuations-dissipation relations. We finally demonstrated that the quantum dynamics of the mirrors admits a stochastic interpretation and we derived the corresponding Langevin equations.

The theory we developed provides a quantum description of the dynamics of the mechanical components in a typical opto-mechanical system, that takes into account the multi-mode nature of the field. It provides an understanding of the fundamental mechanism by which quantum fields, and in particular the zero-point fluctuations that populate the vacuum state, mediate the interaction and thus the exchange of energy between movable components. Given the rapid advances in the miniaturization and cooling techniques, that allows mechanical devices to operate close the quantum level, this effect is promising in view of the development of quantum devices, that could serve as sensors or actuators of quantum motion.

\section{Acknowledgments}
Continuous stimulating discussions with Stephen Barnett, Iacopo Carusotto, Bei Lok Hu, Jen Tsung Hsiang and Kanupriya Sinha are warmly acknowledged. The author acknowledges funding from the Leverhulme Trust Grant No.~ECF-2019-461, and from University of Glasgow via the Lord Kelvin/Adam Smith (LKAS) Leadership Fellowship.

\newpage
\appendix

\section{Field Lagrangian in the instantaneous basis \label{App1}}
We detail here the calculations that lead to the action in Eq.\eqref{Eq:LA}.
For the sake of readability, we rewrite here the Lagrangian of the field:
\begin{equation}
    L_A'[A,q_n] = 
	\frac{1}{2}\int_{q_L(t)}^{q_R(t)}{dx \left[(\partial_t A)^2-(\partial_x A)^2\right]},
	\label{EqA1:LA}
\end{equation}
together with the expansion of the field in the instantaneous basis
\begin{equation}
	A(x,t) = \sum_k {Q_k(t)\varphi_k[x,q_n(t)]},
	\label{EqA1:ModesDecomp}
\end{equation}
where the eigenmodes with vanishing Dirichlet boundary conditions in correspondence to the position of the mirrors have the form:
\begin{equation}
	\varphi_k[x;q_n(t)] \equiv \sqrt{\frac{2}{q_R(t)-q_L(t)}}\,\sin\left[\omega_k(t) (x-q_L(t))\right].
	\label{EqA1:Modes}
\end{equation}
We remind that $\omega_k(t) \equiv k\pi/(q_R(t)-q_L(t))$ is the cavity-length-dependent frequency of the $k-$th modes. We proceed by calculating first the time and spatial derivative of the field $A(x,t)$. In the case of the electromagnetic field, these physically represent the electric field and magnetic fields, respectively:
\begin{align}
    \partial_t A &= \sum_k\left(\dot{Q}_k(t)\varphi_k + Q_k(t)\frac{\partial\varphi_k}{\partial q_R}\dot{q}_R + Q_k(t)\frac{\partial\varphi_k}{\partial q_L}\dot{q}_L\right),\\
    \partial_x A &= \sum_k Q_k(t)\frac{\partial\varphi_k}{\partial x}.
 \end{align}
These are then squared, giving:
\begin{align}
    (\partial_t A)^2 &=\sum_{k,j}\left[\dot{Q}_k\dot{Q}_j\varphi_k \varphi_j + 2\dot{Q}_k Q_j \varphi_k \left(\frac{\partial\varphi_j}{\partial q_R}\dot{q}_R + \frac{\partial\varphi_j}{\partial q_L}\dot{q}_L\right) + Q_k Q_j \left(\frac{\partial\varphi_k}{\partial q_R}\dot{q}_R + \frac{\partial\varphi_k}{\partial q_L}\dot{q}_L\right)\left(\frac{\partial\varphi_j}{\partial q_R}\dot{q}_R + \frac{\partial\varphi_j}{\partial q_L}\dot{q}_L\right) \right],\label{EqA1:DADt2}\\
    (\partial_x A)^2 &=\sum_{k,j}\left({Q}_k{Q}_j\frac{\partial\varphi_k}{\partial x} \frac{\partial\varphi_j}{\partial x}  \right). \label{EqA1:DADx2}
 \end{align}
The Lagrangian of the field in terms of the amplitudes $Q_{\{k\}}$ and the velocities $\dot{Q}_{\{k\}}$ is finally obtained by substituting Eqs.~\eqref{EqA1:DADt2} and \eqref{EqA1:DADx2} into \eqref{EqA1:LA}, and integrating between the positions $q_L$ and $q_R$, respectively of the left and right mirrors:
\begin{equation}
\begin{split}
    L_A &= \frac{1}{2}\sum_{kj}\left\{\dot{Q}_k\dot{Q}_j \left(\int_{q_L}^{q_R}{dx\,\varphi_k\varphi_j}\right) + 2\dot{Q}_k Q_j \left[ \left(\int_{q_L}^{q_R}{dx\,\varphi_k \frac{\partial\varphi_j}{\partial q_R} }\right) \dot{q}_R+\left(\int_{q_L}^{q_R}{dx\,\varphi_k \frac{\partial\varphi_j}{\partial q_L} }\right) \dot{q}_L\right]\right.\\
    &\left. + Q_k Q_j\left[ \left(\int_{q_L}^{q_R}{dx\,\frac{\partial\varphi_k}{\partial q_R} \frac{\partial\varphi_j}{\partial q_R} }\right)\dot{q}_R^2 + 2 \left(\int_{q_L}^{q_R}{dx\,\frac{\partial\varphi_k}{\partial q_R} \frac{\partial\varphi_j}{\partial q_L} }\right)\dot{q}_R \dot{q}_L + \left(\int_{q_L}^{q_R}{dx\,\frac{\partial\varphi_k}{\partial q_L} \frac{\partial\varphi_j}{\partial q_L} }\right)\dot{q}_L^2 \right] \right.\\
    &\left. - Q_k Q_j  \left(\int_{q_L}^{q_R}{dx\,\frac{\partial\varphi_k}{\partial x} \frac{\partial\varphi_j}{\partial x} }\right) \right\}.
\end{split}
\label{Eq:App-LA}
\end{equation}
By using the Eqs.~\eqref{Eq:r_kj}-\eqref{Eq:g_kj} and \eqref{Eq:r_kl-r_jl}-\eqref{Eq:l_kl-r_jl} given in the main text, and noting that
\[\left(\int_{q_L}^{q_R}{dx\,\frac{\partial\varphi_k}{\partial x} \frac{\partial\varphi_j}{\partial x} }\right) = {\omega_k^2 Q_k^2}\,\delta_{k,j},\]
where $\delta_{k,j}$ is the standard Kronecker delta, Eq.~\eqref{Eq:App-LA} reduces to
\begin{equation}
	L_A' = {\left\{\frac{1}{2}\sum_k{[\dot{Q}_k^2 - \omega_k^2(t)Q_k^2]} + \sum_{k,j}\frac{\dot{Q}_k Q_j}{q_R-q_L}\dot{x}_{jk} + \sum_{k,j,\ell}{\frac{Q_k Q_j}{(q_R-q_L)^2} \dot{x}_{j\ell}\dot{x}_{k\ell}}\right\}}.
\end{equation}
This is the Lagrangian that appears in Eq.~\eqref{Eq:LA}.

\section{Influence action \label{App2}}
We detail here the calculations that lead to the influence action for the two mirrors, in the form that is given in Sec.~\ref{Sec:IA} in the main text. We divide the following arguments into two sections. In Sec.~\ref{Sec:App_IA_I}, we calculate the first order term of the perturbative expansion for the influence action, that is Eq.~\eqref{Eq:A1}. In Sec.~\ref{Sec:App_IA_II}, we detail instead the steps that lead to the second order term, that is Eq.~\eqref{Eq:A2}.

\subsection{First order term\label{Sec:App_IA_I}}
We report here again the definition of first order term of the influence action for completeness. This takes the form:
\begin{equation}
    \delta A^{(1)}[q_n, q'_n] = \big<S_{\rm int}[Q_{\{k\}},q_n]\big>_{\rm A} - \big<S_{\rm int}[Q'_{\{k\}},q'_n]\big>_{\rm A}.\label{EqA:A1}
\end{equation}
We calculate Eq.~\eqref{EqA:A1} by using the interaction action $S_{\rm int}[q_n, q'_n]$ given in Eq.~\eqref{Eq:Sint}. For the sake of the arguments discussed in this section, we find it is convenient to break out $S_{\rm int}[q_n, q'_n]$ into the three different contributions:
\begin{equation}
	S_{\rm int}[q_n, q'_n] \equiv S_{\rm int,I}[q_n, q'_n] + S_{\rm int,II}[q_n, q'_n] + S_{\rm int,III}[q_n, q'_n],
\end{equation}
with
\begin{subequations}
\begin{align}
	S_{\rm int,I} &=\int_{t_i}^{t_f}{dt\,}\frac{1}{2}\sum_k{[\omega_{k,0}^2 - \omega_k^2(t)]Q_k^2},\label{Eq:Sint_I}\\
	S_{\rm int,II} &=\int_{t_i}^{t_f}{dt\,}\sum_{k,j}\frac{\dot{Q}_k Q_j}{q_R-q_L}\dot{x}_{jk},\label{Eq:Sint_II}\\
	S_{\rm int,III} &=\int_{t_i}^{t_f}{dt\,}\sum_{k,j,\ell}{\frac{Q_k Q_j}{(q_R-q_L)^2} \dot{x}_{j\ell}\dot{x}_{k\ell}},\label{Eq:Sint_III}
\end{align}
\end{subequations}
In Eq.~\eqref{Eq:Sint_I}, we write $\omega_k^2(t) = \omega_{k,0}^2(1+f(q_\Delta))$, where the function $f(q_\Delta)$ accounts for the variation of the mode frequency induced by a change in the relative distance between the mirrors. By remembering that $g_{kj} = 0 $ for $k=j$, and given that the second order cross-correlations $\big<\dot{Q}_k Q_j\big>=\big<Q_k Q_j\big>=0$ for $k\neq j$ (each mode of the free field is an independent harmonic oscillator), we obtain:
\begin{align}
    \delta A_{\rm I}^{(1)}[ q_n, q'_n] &\equiv \big<S_{\rm int,I}[Q_{\{k\}},q_n]\big>_{\rm A} - \big<S_{\rm int,I}[Q'_{\{k\}},q'_n]\big>_{\rm A} \nonumber\\
    &= \int_{t_i}^{t_f}{dt\,\bigg\{-\frac{1}{2}\sum_k\omega_{k,0}^2\left[\left<Q_k^2\right>_{\rm A}f(q_\Delta) - \left<(Q_k')^2\right>_{\rm A}f(q_\Delta')\right]\bigg\}}\nonumber\\
    & = \int_{t_i}^{t_f}{dt\,\bigg\{-\frac{1}{2}\sum_k\hbar\omega_{k,0}^2\nu_{k}(0)\left[f(q_\Delta) - f(q_\Delta')\right]\bigg\}}\nonumber\\
   & = \int_{t_i}^{t_f}{dt\,\bigg\{-\frac{1}{2}\left(\sum_k\frac{\hbar\omega_{k,0}}{2}z_k\right)\left[f(q_\Delta) - f(q_\Delta')\right]\bigg\}},\label{Eq:A1_I}\\[18pt]
   \delta A_{\rm II}^{(1)}[q_n, q'_n] &= 0,\\[18pt]
       \delta A_{\rm III}^{(1)}[q_n, q'_n] &= \frac{1}{2L_0^2}\int_{t_i}^{t_f}{dt\,\bigg\{\sum_{k,j,l}\left[\left<Q_k Q_j\right>_{\rm A} \dot{x}_{kl}\dot{x}_{jl} - \left<Q_k' Q_j'\right>_{\rm A} \dot{x}_{kl}'\dot{x}_{jl}'\right]\bigg\}}\nonumber\\
   & = \frac{1}{2L_0^2}\int_{t_i}^{t_f}{dt\,\bigg\{\sum_{k,j}\left[\big<Q_k^2\big>_{\rm A} (\dot{x}_{kj})^2 - \big<{Q_k'}^2 \big>_{\rm A} (\dot{x}_{kj}')^2\right]\bigg\}}\nonumber\\
   & = \frac{1}{2L_0^2}\int_{t_i}^{t_f}{dt\,\bigg\{\sum_{k,j}\left(\frac{\hbar z_k}{2\omega_{k,0}}\right)\left[ (\dot{x}_{kj})^2 - ({\dot{x}_{kj}'})^2\right]\bigg\}}. \label{Eq:A1_III}
\end{align}
Here we used the Feynman propagators in Eq.~\eqref{Eq:PropQ}(a-c), together with the noise and dissipation kernels in Eqs.~\eqref{Eq:Nu} and \eqref{Eq:Mu}. We note that the first term, Eq.~\eqref{Eq:A1_I}, diverges, as it is proportional to the energy:
\begin{equation}
	E_{A} = \sum_k\frac{\hbar\omega_{k,0}}{2}z_k.
\end{equation}
This term can be renormalized by subtracting the value it takes in free-space, that is in absence of boundary conditions imposed to the field \cite{birrell1984quantum}. This condition is attained in the limit of infinite separation between the mirrors: $L_0\to\infty$. Note that the physical quantity that can be renormalized is the energy density of the field, which is an intensive quantity, as opposed to the energy, that is an extensive quantity instead and thus implicitly depending on the separation $L_0$. Because of this reason, the renormalization procedure starts by writing the action $\delta A_{\rm I}^{(1)}[ q_n, q'_n]$ in terms of the Lagrangian density:
\begin{equation}
	\delta A_{\rm I}^{(1)}[ q_n, q'_n] = \int_{t_i}^{t_f}{dt\,\int_0^{L_0}dx\left\{-\frac{1}{2}\sum_k\left(z_k\frac{\hbar\omega_{k,0}}{2L_0}\right)\left[f(q_\Delta) - f(q_\Delta')\right]\right\}}.\label{Eq:App2:A1_I}
\end{equation}
We regularize the divergent energy density:
\begin{equation}
	\epsilon_{A} \equiv \frac{E_{A}}{L_0} = \left(\sum_k z_k\frac{\hbar\omega_{k,0}}{2L_0}\right),
\end{equation}
by introducing the frequency cut-off $\sigma^{-1}$ in the sum over all modes:
\begin{equation}
	\epsilon_{A}^{\rm reg} \equiv \left(\sum_k z_k\frac{\hbar\omega_{k,0}}{2L_0} \exp{(-\sigma\omega_{k,0})}\right).
\end{equation}
The renormalized energy density is thus obtained by eliminating the (divergent) value this attains in the $L_0\to\infty$ limit:
\begin{equation}
	\epsilon_{A}^{\rm ren} \equiv \lim_{\sigma\to 0}\left[\epsilon_{A}^{\rm reg} - \lim_{L_0\to \infty}\epsilon_{A}^{\rm ren}\right].
\end{equation}
Let us consider the relevant zero temperature and high temperature limits. In the former case $(z_k\to 1)$, the regularized energy density takes the form:
\begin{align}
	\epsilon_{A,0}^{\rm reg} = \sum_{k=0}^{+\infty}\frac{\hbar\omega_{k,0}}{2L_0}\exp{(-\sigma\omega_{k,0})} = \frac{\hbar}{2\pi\sigma^2} - \frac{\hbar\pi}{24L_0^2} + \mathcal{O}(\sigma^2).
\label{Eq:App2:RegT0}
\end{align}
We recognize in the first term of Eq.~\eqref{Eq:App2:RegT0} the vacuum divergence that need to be subtracted to obtain the renormalized energy density of the field. By proceeding in this way, we get the Casimir energy density
\begin{equation}
	\epsilon_{A,0}^{\rm ren} = - \frac{\hbar\pi}{24L_0^2}.
\end{equation}
The corresponding renormalized action, at low temperature, takes thus the form:
\begin{equation}
\begin{split}
	\delta A_{\rm I,ren}^{(1)}[q_n,q'_n] &\equiv \lim_{\sigma\to 0}\left(\delta A_{\rm I,reg}^{(1)}[q_n,q'_n]-\lim_{L_0\to \infty} \delta A_{\rm I,reg}^{(1)}[q_n,q'_n]\right) \\
	&=  \frac{1}{2} \int_{t_i}^{t_f}{dt \,\left(\frac{\hbar\pi}{24L_0}\right)\left[f(q_\Delta) - f(q_\Delta')\right]} \qquad \text{low temp.}
\end{split}
	\label{Eq:A1_ren_Low}
\end{equation}
The same procedure can be followed in the opposite, high-temperature limit. This limit is attained by first posing the cut-off frequency $\sigma^{-1}$, and then assuming ${k_B T}/{\hbar\sigma^{-1}}\gg 1$. In this regime: $(z_k\to {2k_BT}/{\hbar\omega_{k,0}})$ and the renormalized action takes the form:
\begin{equation}
	\delta A_{\rm I,ren}^{(1)}[q_n,q'_n] = -\frac{1}{2} \int_{t_i}^{t_f}{dt \,\left(\frac{k_B T}{2}\right)\left[f(q_\Delta) - f(q_\Delta')\right]} \qquad \text{high temp.}	
	\label{Eq:A1_ren_High}
\end{equation}

The last step of our calculations entails assuming small oscillations of the mirrors respect to their equilibrium positions $q_{n,0}$, such that: $q_n(t) = q_{n,0} + \delta q_{n}(t)$, where $\delta q_n(t)$ denotes fluctuations of the mirrors' positions. This allows us to expand the function $f(q_\Delta)$ in powers of the parameter $\delta q_n/L_0$. Up to second order, this gives:
\begin{equation}
	f(q_\Delta) = 1/(1+\delta q_\Delta/L_0)^2 \approx 1-2 \frac{\delta q_\Delta}{L_0} + 3 \left(\frac{\delta q_\Delta}{L_0}\right)^2.
\end{equation}
By using this expansion into Eq.~\eqref{Eq:App2:A1_I} (or into the renormalized versions \eqref{Eq:A1_ren_Low} and \eqref{Eq:A1_ren_High}), we recognize that the term of first order in the position fluctuations accounts for the static Casimir force, while the term of second order represents a correction to the trapping experienced by the relative distance dof of the two mirrors. At zero temperature, these take the form:
\begin{subequations}
\begin{align}
	\delta A_{\rm I,ren}^{(1),\delta}[\delta q_\Delta,\delta q'_\Delta] 	&=  - \int_{t_i}^{t_f}{dt \,\left(\frac{\hbar\pi}{24L_0^2}\right)\left(\delta q_\Delta - \delta q_\Delta'\right)}, \qquad\;\; \text{low temp.}\\
	 \delta A_{\rm I,ren}^{(1),\delta^2}[\delta q_\Delta,\delta q'_\Delta] &=  \int_{t_i}^{t_f}{dt \,\left(\frac{\hbar\pi}{16L_0^3}\right)\left[(\delta q_\Delta^2) - (\delta q_\Delta')^2\right]}  \qquad \text{low temp.}
\end{align}
\end{subequations}
while in the high temperature limit:
\begin{subequations}
\begin{align}
	\delta A_{\rm I,ren}^{(1),\delta}[\delta q_\Delta,\delta q'_\Delta] 	&=  \int_{t_i}^{t_f}{dt \,\left(\frac{K_B T}{2L_0}\right)\left(\delta q_\Delta - \delta q_\Delta'\right)}, \qquad \qquad\;\;\;\text{high temp.}\\
	 \delta A_{\rm I,ren}^{(1),\delta^2}[\delta q_\Delta,\delta q'_\Delta] &=  - \int_{t_i}^{t_f}{dt \,\left(\frac{3K_B T}{4L_0^2}\right)\left[(\delta q_\Delta^2) - (\delta q_\Delta')^2\right]} \qquad \text{high temp.} 
\end{align}
\end{subequations}

We do not discuss here the last, non-zero, $\delta A_{\rm III}^{(1)}[q_n, q'_n]$ term in Eq.~\eqref{Eq:A1_III}. This term cancels out from the final expression of the influence action, as the same quantity, with opposite sign, appears in the second order term $\delta A^{(2)}[\delta q_n,\delta q_n']$, whose calculation is illustrated in the next section.

\subsection{Second order term\label{Sec:App_IA_II}}

We detail here the calculation of the second order contribution to the influence action, that is Eq.~\eqref{Eq:A2} in the main text. For ease of the following arguments, it is convenient to label the different terms that compose it as follows:
\begin{equation}
	\delta A^{(2)}[ q_n, q'_n ] = \delta A_{a}^{(2)}[q_n] + \delta A_{b}^{(2)}[q'_n ] + \delta A_{c}^{(2)}[q_n,q'_n ],
\end{equation}
with
\begin{subequations}
\begin{align}
    \delta A_{a}^{(2)}[q_n] &= \frac{i}{2\hbar}\left\{ \big<{S_{\rm int}^2}[Q_{\{k\}},q_n]\big>_{\rm A} - \big<S_{\rm int}[Q_{\{k\}},q_n]\big>_{\rm A}^2 \right\},\\
    \delta A_{b}^{(2)}[q'_n ] &=\frac{i}{2\hbar}\left\{ \big<{S_{\rm int}^2}[Q'_{\{k\}},q'_n]\big>_{\rm A} - \big<S_{\rm int}[Q'_{\{k\}},q'_n]\big>_{\rm A}^2 \right\},\\
    \delta A_{c}^{(2)}[q_n,q'_n ] &= -\frac{i}{\hbar}\left\{\big<S_{\rm int}[Q_{\{k\}},q_n]S_{\rm int}[Q'_{\{k\}},q'_n]\big>_{\rm A} - \big<S_{\rm int}[q_n,Q_{\{k\}}]\big>_{\rm A}\big<S_{\rm int}[Q'_{\{k\}},q'_n]\big>_{\rm A}\right\}.
\end{align}
\label{Eq:App2:IA2-abc}
\end{subequations}
Since the unperturbed dynamics of the field is quadratic, the fourth-order correlators that arise by substituting $S_{\rm int}[Q_{\{k\}},q_n]$ into Eqs.~\eqref{Eq:App2:IA2-abc}(a-c), can be decomposed in terms of second-order correlators only. These involve both the amplitudes $Q_{\{k\}}$ of the field modes, and the velocities $\dot{Q}_{\{k\}}$. By using Eqs.~\eqref{Eq:PropQ}(a-c), we obtain the following Feynman propagators:
\begin{subequations}
\begin{align}
	&\mathcal{D}_{\dot{Q}_k Q_k}(s_1-s_2) \equiv \frac{d}{ds_1}\mathcal{D}_{Q_k Q_k}(s_1-s_2) =-i\hbar[-\dot{\mu}_{k}(s_1-s_2){\rm sgn}(s_1-s_2)+i\dot{\nu}_{k}(s_1-s_2)],\\
	&\mathcal{D}_{\dot{Q}_k \dot{Q}_k}(s_1-s_2) \equiv \frac{d^2}{ds_1ds_2}\mathcal{D}_{Q_k Q_k}(s_1-s_2) =i\hbar[-\ddot{\mu}_{k}(s_1-s_2){\rm sgn}(s_1-s_2)+i\ddot{\nu}_{k}(s_1-s_2)-2\dot{\mu}_k(0)\delta(s_1-s_2)],\\
	&\mathcal{D}_{\dot{Q}'_k Q'_k}(s_1-s_2) \equiv \frac{d}{ds_1}\mathcal{D}_{Q'_k Q'_k}(s_1-s_2)=-i\hbar[\dot{\mu}_{k}(s_1-s_2){\rm sgn}(s_1-s_2)+i\dot{\nu}_{k}(s_1-s_2)],\\
	&\mathcal{D}_{\dot{Q}'_k \dot{Q}'_k}(s_1-s_2) \equiv \frac{d^2}{ds_1ds_2}\mathcal{D}_{Q'_k Q'_k}(s_1-s_2)=i\hbar[\ddot{\mu}_{k}(s_1-s_2){\rm sgn}(s_1-s_2)+i\ddot{\nu}_{k}(s_1-s_2)+2\dot{\mu}_k(0)\delta(s_1-s_2)],\\
	&\mathcal{D}_{\dot{Q}'_k Q_k}(s_1-s_2) \equiv \frac{d}{ds_1}\mathcal{D}_{Q'_k Q_k}(s_1-s_2) =-i\hbar[\dot{\mu}_{k}(s_1-s_2)-i\dot{\nu}_{k}(s_1-s_2)],\\
	&\mathcal{D}_{{Q}'_k \dot{Q}_k}(s_1-s_2) \equiv \frac{d}{ds_2}\mathcal{D}_{Q'_k Q_k}(s_1-s_2) =i\hbar[\dot{\mu}_{k}(s_1-s_2)-i\dot{\nu}_{k}(s_1-s_2)],\\
	&\mathcal{D}_{\dot{Q}'_k \dot{Q}_k}(s_1-s_2) \equiv \frac{d^2}{ds_1 ds_2}\mathcal{D}_{Q'_k Q_k}(s_1-s_2) =i\hbar[\ddot{\mu}_{k}(s_1-s_2)-i\ddot{\nu}_{k}(s_1-s_2)].
\end{align}
\label{Eq:DerPropagators}
\end{subequations}
Since we are interested in calculating the influence action up to second-order in the fluctuations $\delta q_n/L_0$ respect to the equilibrium positions of the mirrors, we need to insert into Eqs.~\eqref{Eq:App2:IA2-abc}(a-c) only the first order term of the action $S_{\rm int}[Q_{\{k\}},q_n]$. This terms, we call it $S_{\rm int}^{(1)}$, can be composed into two contributions:
\begin{equation}
S^{(1)}_{\rm int}[Q_{\{k\}},\delta q_n] = S^{(1)}_{\rm int, I}[Q_{\{k\}},\delta q_n] + S^{(1)}_{\rm int, II}[Q_{\{k\}},\delta q_n],  \end{equation}
having the form:
\begin{subequations}
\begin{align}
    S^{(1)}_{\rm int, I}[Q_{\{k\}},\delta q_n] &= \int_{t_i}^{t_f}{dt\, \frac{1}{L_0}\sum_k \omega_{k,0}^2 Q_k^2(t) (\delta q_R(t) - \delta q_L(t))},\\
    S^{(1)}_{\rm int, II}[Q_{\{k\}},\delta q_n] &= \int_{t_i}^{t_f}{dt\, \frac{1}{L_0}\sum_{k,j}\dot{Q}_k(t) Q_j(t) \delta \dot{x}_{jk}(t)}.
\end{align}
\end{subequations}
The final expression of $\delta A^{(2)}$ is thus composed by six different terms, that we denote as $\delta A^{(2)}_{i,{\rm I}}$, $\delta A^{(2)}_{i,{\rm II}}$, with $i=a,b,c$. These are obtained by evaluating Eqs.~\eqref{Eq:App2:IA2-abc}(a-c), respectively with $S^{(1)}_{\rm int, I}[Q_{\{k\}},\delta q_n]$ and $S^{(1)}_{\rm int, II}[Q_{\{k\}},\delta q_n]$. Note that, in the same equations, the terms  mixing $S^{(1)}_{\rm int, I}$ and $S^{(1)}_{\rm int, II}$ are identically zero.

In particular we write:
\begin{equation}
	\delta A_a^{(2)}[\delta q_n] = \delta A_{\rm a,I}^{(2)}[\delta q_n] + \delta A_{\rm a,II}^{(2)}[\delta q_n],
\end{equation}
with
\begin{align}
    \delta A_{\rm a,I}^{(2)}&[\delta q_n] = \frac{i}{2\hbar}\left\{ \big<{S^{(1)}_{\rm int, I}}^2[Q_{\{k\}},\delta q_n]\big>_{\rm A} - \big<S_{\rm int, II}^{(1)}[Q_{\{k\}},\delta q_n]\big>_{\rm A}^2 \right\}\nonumber\\
        &=\int_{t_i}^{t_f} ds_1\, \int_{t_i}^{t_f} ds_2 \, \delta q_\Delta(s_1) \delta q_\Delta(s_2)  \left\{\frac{i}{2\hbar L_0^2}\sum_{k,j}{\omega_{k,0}^2\omega_{j,0}^2}\left[\big<Q_k^2(s_1)Q_j^2(s_2)\big>_{\rm A} - \big<Q_k^2(s_1)\big>_{\rm A}\big<Q_j^2(s_2)\big>_{\rm A}\right]   \right\}\nonumber\\
        &=\int_{t_i}^{t_f} ds_1\, \int_{t_i}^{t_f} ds_2 \, \delta q_\Delta(s_1) \delta q_\Delta(s_2)  \left\{\frac{i}{\hbar L_0^2}\sum_{k}{\omega_{k,0}^4}\left[\big<Q_k(s_1)Q_k(s_2)\big>_{\rm A}^2\right]   \right\},\\[18pt]
        \delta A_{\rm a,II}^{(2)}&[\delta q_n] = \frac{i}{2\hbar}\left\{ \big<{S^{(1)}_{\rm int, II}}^2[Q_{\{k\}},\delta q_n]\big>_{\rm A} - \big<S_{\rm int, II}^{(1)}[Q_{\{k\}},\delta q_n]\big>_{\rm A}^2 \right\}\nonumber\\
        &=\int_{t_i}^{t_f} ds_1\, \int_{t_i}^{t_f} ds_2 \,  \Bigg\{\frac{i}{2\hbar L_0^2}\sum_{k,j}\sum_{p,q} \delta \dot{x}_{jk}(s_1) \dot{x}_{qp}(s_2)\left[\big<\dot{Q}_k(s_1){Q}_j(s_1)\dot{Q}_p(s_2){Q}_q(s_2)\big>_{\rm A} \right.\nonumber \\
       &\hspace{3cm}-\left.  \big<\dot{Q}_k(s_1){Q}_j(s_1)\big>_{\rm A}\big<\dot{Q}_p(s_2){Q}_q(s_2)\big>_{\rm A}\right]   \Bigg\}\nonumber\\
        &=\int_{t_i}^{t_f} ds_1\, \int_{t_i}^{t_f} ds_2 \,  \Bigg\{\frac{i}{2\hbar L_0^2}\sum_{k,j}  \left[\delta\dot{x}_{jk}(s_1) \dot{x}_{jk}(s_2)\big<\dot{Q}_k(s_1)\dot{Q}_k(s_1)\big>_{\rm A}\big<{Q}_j(s_2){Q}_j(s_2)\big>_{\rm A} \right.\nonumber\\
       &\hspace{3cm}+\left. \dot{x}_{jk}(s_1) \dot{x}_{kj}(s_2)\big<\dot{Q}_k(s_1){Q}_k(s_1)\big>_{\rm A}\big<{Q}_j(s_2)\dot{Q}_j(s_2)\big>_{\rm A} \right]  \Bigg\},
\end{align}
and
\begin{equation}
	\delta A_b^{(2)}[\delta q_n'] = \delta A_{\rm b,I}^{(2)}[\delta q_n'] + \delta A_{\rm b,II}^{(2)}[\delta q_n'],
\end{equation}
with
\begin{align}
&\begin{aligned}
    \delta A_{\rm b,I}^{(2)}[\delta q_n'] &= \frac{i}{2\hbar}\left\{ \big<{S^{(1)}_{\rm int, I}}^2[Q_{\{k\}}',\delta q_n']\big>_{\rm A} - \big<S_{\rm int, II}^{(1)}[Q_{\{k\}}',\delta q_n']\big>_{\rm A}^2 \right\}\\
        &=\int_{t_i}^{t_f} ds_1\, \int_{t_i}^{t_f} ds_2 \, \delta q_\Delta'(s_1) \delta q_\Delta'(s_2)  \left\{\frac{i}{\hbar L_0^2}\sum_{k}{\omega_{k,0}^4}\left[\big<Q_k'(s_1)Q_k'(s_2)\big>_{\rm A}^2\right]   \right\},
\end{aligned}\\[18pt]
&\begin{aligned}        
        \delta A_{\rm b,II}^{(2)}[\delta q_n'] &= \frac{i}{2\hbar}\left\{ \big<{S^{(1)}_{\rm int, II}}^2[Q_{\{k\}}',\delta q_n']\big>_{\rm A} - \big<S_{\rm int, II}^{(1)}[Q_{\{k\}}',\delta q_n']\big>_{\rm A}^2 \right\}\\
        &=\int_{t_i}^{t_f} ds_1\, \int_{t_i}^{t_f} ds_2 \,  \Bigg\{\frac{i}{2\hbar L_0^2}\sum_{k,j}  \left[\delta\dot{x}_{jk}'(s_1) \dot{x}_{jk}'(s_2)\big<\dot{Q}_k'(s_1)\dot{Q}_k'(s_1)\big>_{\rm A}\big<{Q}_j'(s_2){Q}_j'(s_2)\big>_{\rm A} \right.\\
       &\hspace{3cm}+\left. \dot{x}_{jk}'(s_1) \dot{x}_{kj}'(s_2)\big<\dot{Q}_k'(s_1){Q}_k'(s_1)\big>_{\rm A}\big<{Q}_j'(s_2)\dot{Q}_j'(s_2)\big>_{\rm A} \right]  \Bigg\},
\end{aligned}
\end{align}
and finally
\begin{equation}
	\delta A_c^{(2)}[\delta q_n,\delta q_n'] = \delta A_{\rm c,I}^{(2)}[\delta q_n,\delta q_n'] + \delta A_{\rm c,II}^{(2)}[\delta q_n,\delta q_n'],
\end{equation}
with
\begin{align}
    \delta A_{\rm c,I}^{(2)}&[\delta q_n,\delta q'_n ] = \frac{i}{2\hbar}\left\{ \big<{S^{(1)}_{\rm int, I}}[Q_{\{k\}},\delta q_n]{S^{(1)}_{\rm int, I}}[Q_{\{k\}}',\delta q_n']\big>_{\rm A} - \big<S_{\rm int, I}^{(1)}[\delta q_n,Q_{\{k\}}]\big>_{\rm A}\big<S_{\rm int, I}^{(1)}[\delta q_n',Q_{\{k\}}']\big>_{\rm A} \right\},\nonumber\\
       &=-\int_{t_i}^{t_f} ds_1\, \int_{t_i}^{t_f} ds_2 \, \delta q_\Delta(s_1) \delta q_\Delta'(s_2)  \left\{\frac{i}{\hbar L_0^2}\sum_{k,j}{\omega_{k,0}^2\omega_{j,0}^2}\left[\big<Q_k^2(s_1){Q_j'}^2(s_2)\big> - \big<Q_k^2(s_1)\right>\left<{Q_j'}^2(s_2)\big>\right] \right\}\nonumber\\
        &=-\int_{t_i}^{t_f} ds_1\, \int_{t_i}^{t_f} ds_2 \, \delta q_\Delta(s_1) \delta q_\Delta'(s_2)  \left\{\frac{2i}{\hbar L_0^2}\sum_{k}{\omega_{k,0}^4}\left[\big<Q_k(s_1)Q_k'(s_2)\big>^2\right]   \right\},\\[18pt]
      \delta A_{\rm c,II}^{(2)}&[\delta q_n,\delta q'_n ] = \frac{i}{2\hbar}\left\{ \big<{S^{(1)}_{\rm int, II}}[Q_{\{k\}},\delta q_n]{S^{(1)}_{\rm int, II}}[Q_{\{k\}}',\delta q_n']\big>_{\rm A} - \big<S_{\rm int, II}^{(1)}[Q_{\{k\}},\delta q_n]\big>_{\rm A}\big<S_{\rm int, II}^{(1)}[Q_{\{k\}}',\delta q_n']\big>_{\rm A} \right\}\nonumber\\
      &=-\int_{t_i}^{t_f} ds_1\, \int_{t_i}^{t_f} ds_2 \,  \Bigg\{\frac{i}{\hbar L_0^2}\sum_{k,j}\sum_{p,q} \delta \dot{x}_{jk}(s_1) \dot{x}_{qp}'(s_2)\Bigg[\big<\dot{Q}_k(s_1){Q}_j(s_1)\dot{Q}_p'(s_2){Q}_q'(s_2)\big> \nonumber\\
       &\hspace{3cm}- \big<\dot{Q}_k(s_1){Q}_j(s_1)\big>\big<\dot{Q}_p'(s_2){Q}_q'(s_2)\big>\Bigg]   \Bigg\}\nonumber\\
        &=-\int_{t_i}^{t_f} ds_1\, \int_{t_i}^{t_f} ds_2 \,  \Bigg\{\frac{i}{\hbar L_0^2}\sum_{k,j}  \Bigg[\delta\dot{x}_{jk}(s_1) \dot{x}_{jk}'(s_2)\big<\dot{Q}_k(s_1)\dot{Q}_k'(s_1)\big>\big<{Q}_j(s_2){Q}_j'(s_2)\big>\nonumber\\
       &\hspace{3cm}+\dot{x}_{jk}(s_1) \dot{x}_{kj}'(s_2)\big<\dot{Q}_k(s_1){Q}_k'(s_1)\big>\big<{Q}_j(s_2)\dot{Q}_j'(s_2)\big> \Bigg]   \Bigg\}.       
\end{align}

By combining $\delta A_{\rm a,I}^{(2)}$, $\delta A_{\rm b,I}^{(2)}$ and $\delta A_{\rm c,I}^{(2)}$, we obtain:
\begin{align}
		\delta A_{\rm I}^{(2)}[q_n,q'_n] &= \delta A_{\rm a,I}^{(2)} + \delta A_{\rm b,I}^{(2)} + \delta A_{\rm c,I}^{(2)}\nonumber\\
		&= \hbar\left\{\int_{t_i}^{t_f} ds_1 \int_{t_i}^{s_1} ds_2  \left[-i\delta{q}_\Delta^{(-)}(s_1)\bar{N}(s_1-s_2)\delta{q}_\Delta^{(-)}(s_2)+\delta{q}_\Delta^{(-)}(s_1)\bar{M}(s_1-s_2)\delta{q}_\Delta^{(+)}(s_2)\right]\right\}
		\label{Eq:A2_Delta_Int}.
\end{align}
Here we defined the combination of the forward- and backward-in-time histories $\delta q_\Delta^{(\pm)} \equiv (\delta q_\Delta \pm \delta q_\Delta')$, together with the noise and dissipation kernels:
\begin{align}
    \bar{N}(t) & = \sum_k \frac{\omega_{k,0}^2}{4L_0^2}\bar{N}_k(t), &	\bar{M}(t) & = \sum_k \frac{\omega_{k,0}^2}{4L_0^2}\bar{M}_k(t),\label{Eq:Kernels_bar}
\end{align}
with 
\begin{align*}
	\bar{N}_k(t) & = \nu_+(t;k,k) + \nu_-(t;k,k), &
\bar{M}_k(t) & = \mu_+(t;k,k),
\end{align*}
and
\begin{align*}
 \nu_\pm(t;k,j) &\equiv  - (z_k z_j \pm 1) \cos[(\omega_{k,0}\pm\omega_{j,0})t], &
 \mu_\pm(t;k,j) &\equiv  (z_k \pm z_j ) \sin[(\omega_{k,0}\pm\omega_{j,0})t].
\end{align*}

By combining $\delta A_{\rm a,II}^{(2)}$, $\delta A_{\rm b,II}^{(2)}$ and $\delta A_{\rm c,II}^{(2)}$ instead, we obtain:
\begin{equation}
    \begin{split}
       \delta A_{\rm II}^{(2)}&[q_n,q'_n] = \delta A_{\rm a,II}^{(2)} + \delta A_{\rm b,II}^{(2)} + \delta A_{\rm c,II}^{(2)}\\
       &= \frac{\hbar}{8L_0^2}\Bigg\{\int_{t_i}^{t}ds_1 \int_{t_i}^{s_1} ds_2\sum_{kj}\Bigg[ -i\,\nu_{20}^{kj}(s_1-s_2)\left(\delta\dot{x}_{jk}(s_1)-\delta\dot{x}_{jk}'(s_1)\right)\left(\delta\dot{x}_{jk}(s_2)-\delta\dot{x}_{jk}'(s_2)\right)\\
        &\hspace{3cm}+\mu_{20}^{kj}(s_1-s_2)\left(\delta\dot{x}_{jk}(s_1)-\delta\dot{x}_{jk}'(s_1)\right)\left(\delta\dot{x}_{jk}(s_2)-\delta\dot{x}_{jk}'(s_2)\right)\Bigg]\Bigg\}\\
        &+\frac{\hbar}{8L_0^2}\Bigg\{\int_{t_i}^{t}ds_1 \int_{t_i}^{s_1} ds_2\sum_{kj}\Bigg[ -i\,\nu_{11}^{kj}(s_1-s_2)\left(\delta\dot{x}_{jk}(s_1)-\delta\dot{x}_{jk}'(s_1)\right)\left(\delta\dot{x}_{jk}(s_2)-\delta\dot{x}_{jk}'(s_2)\right)\\
        &\hspace{3cm}+\mu_{11}^{kj}(s_1-s_2)\left(\delta\dot{x}_{jk}(s_1)-\delta\dot{x}_{jk}'(s_1)\right)\left(\delta\dot{x}_{jk}(s_2)-\delta\dot{x}_{jk}'(s_2)\right)\Bigg]\Bigg\}\\
        &+\frac{i\hbar}{8L_0^2}\Bigg\{\int_{t_i}^{t_f}ds_1 \int_{t_i}^{t_f} ds_2\sum_{kj}\Bigg[i\dot{\mu}_{k}(0){\nu}_{k}(0)\delta(s_1-s_2)\left(\delta\dot{x}_{jk}'(s_2)-\delta\dot{x}_{jk}'(s_2)-\delta\dot{x}_{jk}(s_2)-\delta\dot{x}_{jk}(s_2)\right)\Bigg]\Bigg\},
    \end{split}
    \label{Eq:App2-A2II}
\end{equation}
where we defined
\begin{align*}
	\nu_{11}^{kj}(t) &\equiv \dot{\nu}_k(t)\dot{\nu}_j(s) - \dot{\mu}_k(t)\dot{\mu}_j(t) = -\nu_+(t;k,j) + \nu_-(t;k,j) ,\\
	\mu_{11}^{kj}(t) &\equiv \dot{\nu}_k(t)\dot{\mu}_j(t) + \dot{\mu}_k(t)\dot{\nu}_j(t) = \mu_+(t;k,j) + \mu_-(t;k,j),\\
	\nu_{20}^{kj}(t) &\equiv \ddot{\nu}_k(t)\nu_j(t) - \ddot{\mu}_k(t)\mu_j(t) = \frac{\omega_{k,0}}{\omega_{j,0}}\left[-\nu_+(t;k,j) - \nu_-(t;k,j)\right],\\
	\mu_{20}^{kj}(t) &\equiv \ddot{\mu}_k(t)\nu_j(t) + \ddot{\nu}_k(t)\mu_j(t) = \frac{\omega_{k,0}}{\omega_{j,0}}\left[\mu_+(t;k,j) - \mu_-(t;k,j)\right].
\end{align*}
Eq.~\eqref{Eq:App2-A2II} can be further recast in the form:
\begin{equation}
    \begin{split}
        \delta A_{\rm II}^{(2)}[\delta q_n,\delta q_n'] &= \frac{\hbar}{L_0^2}\Bigg\{\int_{t_i}^{t_f}ds_1 \int_{t_i}^{s_1} ds_2\sum_{\substack{kj}}\Bigg[-i \delta\dot{x}_{jk}^{(-)}(s_1)\tilde{N}_{kj}(s_1 - s_2)\delta\dot{x}_{jk}^{(-)}(s_2)\\
        &+\delta\dot{x}_{jk}^{(-)}(s_1)\tilde{M}_{kj}(s_1 - s_2)\delta\dot{x}_{jk}^{(+)}(s_2)\Bigg]-\frac{1}{2L_0^2}\left\{\int_{t_i}^{t_f}ds_1 \sum_{kj}\left(\frac{\hbar z_k}{2\omega_{k,0}}\right)\left[{\delta\dot{x}_{kj}}^2(s_1)-{\delta\dot{x}_{kj}'}^2(s_1)\right]\right\},
    \end{split}
    \label{Eq:App2-A2II-2}
\end{equation}
where we defined the kernels
\begin{align}
	\tilde{N}_{k,j}(t) &= \tilde{N}_{k,j}^{(+)}(t) + \tilde{N}_{k,j}^{(-)}(t), &
	\tilde{M}_{k,j}(t) &= \tilde{M}_{k,j}^{(+)}(t) + \tilde{M}_{k,j}^{(-)}(t),
\end{align}
with 
\begin{align*}
	\tilde{N}_{k,j}^{(\pm)}(t) &= \frac{(\omega_{k,0}\mp\omega_{j,0})^2}{16\omega_{k,0}\omega_{j,0}} \nu_\pm(t;k,j), & 
	\tilde{M}_{k,j}^{(\pm)}(t) &= \frac{(\omega_{k,0}\mp\omega_{j,0})^2}{16\omega_{k,0}\omega_{j,0}} \mu_\pm(t;k,j).
\end{align*}
As anticipated in the previous section, the last term in Eq.~\eqref{Eq:App2-A2II-2} is equal to and opposite in sign respect to the first order contribution $\delta A_{\rm III}^{(1)}[q_n,q_n']$, Eq.~\eqref{Eq:A1_III}, so that the two cancel out.

By using the definition:
\[x_{jk} = r_{jk} q_R + l_{jk} q_L = g_{jk} \left(q_L-(-1)^{k+j}q_R\right),\]
it results that
\begin{align*}
    x_{jk} &= g_{jk} (q_L - q_R) = - g_{jk} q_\Delta, & \text{for}\; j+k &= {\rm even},\\
    x_{jk} &= g_{jk} (q_L + q_R) = g_{jk} q_\Sigma, & \text{for}\; j+k &= {\rm odd}.
\end{align*}
This allows us to recognize in Eq.\eqref{Eq:App2-A2II-2} two different contributions, respectively for the mutual distance between the mirrors and their CM:
\begin{equation}
    \delta A_{\rm II}^{(2)}[\delta q_n,\delta q_n'] = \delta A_{\rm II,\Delta}^{(2)}[\delta q_\Delta,\delta q_\Delta'] + \delta A_{\rm II,\Sigma}^{(2)}[\delta q_\Sigma,\delta q_\Sigma'],
\end{equation}
with
\begin{subequations}
\begin{align}
    \delta &A_{\rm II,\Delta}^{(2)}[\delta \dot{q}_\Delta,\delta \dot{q}_\Delta'] = \hbar\int_{t_i}^{t_f}ds_1 \int_{t_i}^{s_1} ds_2\Bigg[-i \delta\dot{q}_{\Delta}^{(-)}(s_1)N_\Delta(s_1 - s_2)\delta\dot{q}_{\Delta}^{(-)}(s_2)+\delta\dot{q}_{\Delta}^{(-)}(s_1)M_\Delta(s_1 - s_2)\delta\dot{q}_{\Delta}^{(+)}(s_2)\Bigg],\\
    \delta &A_{\rm II,\Sigma}^{(2)}[\delta \dot{q}_\Sigma,\delta \dot{q}_\Sigma'] = \hbar\int_{t_i}^{t_f}ds_1 \int_{t_i}^{s_1} ds_2\Bigg[-i \delta\dot{q}_{\Sigma}^{(-)}(s_1)N_\Sigma(s_1 - s_2)\delta\dot{q}_{\Sigma}^{(-)}(s_2)+\delta\dot{q}_{\Sigma}^{(-)}(s_1)M_\Sigma(s_1 - s_2)\delta\dot{q}_{\Sigma}^{(+)}(s_2)\Bigg].
\end{align}
\end{subequations}
Here we defined the kernels
\begin{align}
    N_{\Delta}(t) &= {\sum_{\substack{kj}}}''\frac{\omega_{k,0}\omega_{j,0}}{4L_0^2} N_{kj}(t), & M_{\Delta}(t) &= {\sum_{\substack{kj}}}''\frac{\omega_{k,0}\omega_{j,0}}{4L_0^2} M_{kj}(t),\label{Eq:Kernels_Delta_App}\\
    N_{\Sigma}(t) &= {\sum_{\substack{kj}}}'\frac{\omega_{k,0}\omega_{j,0}}{4L_0^2} N_{kj}(t),  & M_{\Sigma}(t)&= {\sum_{\substack{kj}}}'\frac{\omega_{k,0}\omega_{j,0}}{4L_0^2} M_{kj}(t).\label{Eq:Kernels_Sigma_App}
\end{align}
with
\begin{align*}
	N_{k,j}(t) &= N_{k,j}^{(+)}(t) + N_{k,j}^{(-)}(t), &
	M_{k,j}(t) &= M_{k,j}^{(+)}(t) + M_{k,j}^{(-)}(t),
\end{align*}
and
\begin{subequations}
\begin{align*}
	N_{k,j}^{(\pm)}(t) &= \frac{1}{(\omega_{k,0}\pm\omega_{j,0})^2} \nu_\pm(t;k,j), & 
	M_{k,j}^{(\pm)}(t) &= \frac{1}{(\omega_{k,0}\pm\omega_{j,0})^2} \mu_\pm(t;k,j)
\end{align*}
\end{subequations}
In Eqs.~\eqref{Eq:Kernels_Delta_App}, double primed sums indicate sums over couples of cavity modes $k,j$, such that $k+j$ is even while, in Eq.~\eqref{Eq:Kernels_Sigma_App}, single primed sums indicate sums over modes such that $k+j$ is odd. This result shows that the dofs of the relative distance between the mirrors and of the CM interact with an environment that is composed by different combinations of field modes. 

\bibliography{TwoMirrors.bib}
\end{document}